\documentclass[journal]{IEEEtran}
\usepackage{graphicx}
\usepackage{amssymb,amsmath,amsthm}
\usepackage[ruled,vlined,linesnumbered,lined,boxed]{algorithm2e}
\usepackage{xcolor}
\usepackage{booktabs}
\usepackage{url}
\usepackage[shortlabels]{enumitem}
\usepackage[latin1]{inputenc}

\DeclareMathOperator{\var}{var} 
\usepackage[caption=false,font=footnotesize]{subfig}
\newcommand{\e}{\mathbf{e}}
\newcommand{\commat}{\bo K_\pdim}  % COMMUTATION matrix 
\newcommand{\tabasco}{\textsc{Tabasco}{}}
\newcommand{\SSCMshape}{{\hat{\boldsymbol \Lambda}}}

\newcounter{ctheorem}
\newtheorem{theorem}[ctheorem]{Theorem}

\newcounter{clemma}
\newtheorem{lemma}[clemma]{Lemma}

\theoremstyle{definition}

\usepackage{hyperref}
\usepackage{tikz}
\usepackage{pgfplots}
\pgfplotsset{compat=1.13}
\newlength\fwidth

\newcommand{\Sym}[1]{\mathbb{R}_{\mathrm{Sym}}^{#1 \times #1}}
\newcommand{\SymC}[1]{\mathbb{C}_{\mathrm{Sym}}^{#1 \times #1}}
\newcommand{\E}{\mathbb{E}}                  % expectation
\newcommand{\bo}[1]{\mathbf{#1}}              % boldface math 
\newcommand{\bom}[1]{\boldsymbol{#1}}    % boldface math (for greek letters)
\newcommand{\ve}{\mathrm{vec}}
\newcommand{\be}{\beta}
\newcommand{\ka}{\kappa}
\newcommand{\pdim}{p}
\newcommand{\bmu}{\bom{\mu}}
\newcommand{\La}{{\boldsymbol \Lambda}}

\renewcommand{\d}{\mathbf{d}}
\newcommand{\etatwo}{\vartheta_{\W}}
\newcommand{\etatwohat}{\hat \vartheta_{\W}}

\newcommand{\hop}{\mathsf{H}}        % Hermitian transpose 

\newcommand{\beq}{\begin{equation}}
\newcommand{\eeq}{\end{equation}}
\newcommand{\bmat}{\begin{pmatrix}}
\newcommand{\emat}{\end{pmatrix}}
\newcommand{\R}{\mathbb{R}}

\newcommand{\shape}{{\boldsymbol \Lambda}}

\renewcommand{\S}{{\bo S}} % SAMPLE COVARIANCE MATRIX
\newcommand{\C}{\mathbb{C}} %  field of complex numbers
\newcommand{\W}{\mathbf{W}}
\newcommand{\V}{\mathbf{V}} 
\newcommand{\A}{\mathbf{A}} 
\newcommand{\B}{\mathbf{B}} 

\newcommand{\x}{\bo x}
\renewcommand{\v}{\bo v}
\newcommand{\w}{\mathbf{w}}

\renewcommand{\a}{{\bo a}}
\newcommand{\Fr}{\mathrm{F}}

\newcommand{\D}{\mathbf{D}}
\newcommand{\M}{\bom \Sigma}
\newcommand{\MSE}{\mathrm{MSE}}

\DeclareMathOperator{\tr}{tr}
\DeclareMathOperator{\Tr}{tr}
\DeclareMathOperator{\diag}{diag}
\DeclareMathOperator{\cov}{cov} 
\usepackage{color}

\begin{document}

\title{Regularized Tapered Sample Covariance Matrix}
\author{
Esa~Ollila,~\IEEEmembership{Senior member,~IEEE}
and
Arnaud Breloy%~\IEEEmembership{Member,~IEEE}
\thanks{
E. Ollila is with the Department of Signal Processing and Acoustics, Aalto University, P.O. Box 15400, FI-0007 Aalto, Finland. 

Arnaud Breloy is with the LEME (EA4416), University Paris Nanterrre, 92410 Ville-d'Avray, France.

The code used in the presented experiments is available in the ${\rm Matlab}$ toolbox https://github.com/esollila/Tabasco.
}% <-this % stops a space
%\thanks{Manuscript received November xx, 2021; revised xx, 202X.}
}

%\markboth{IEEE TRANSACTIONS ON SIGNAL PROCESSING,~Vol.~xxx, No.~x, 2021}%
%{Ollila \MakeLowercase{\textit{et al.}}: \otsikko}




\maketitle

\begin{abstract} 
Covariance matrix tapers have a long history in signal processing and related fields.  
Examples of applications include autoregressive models (promoting a banded structure) or beamforming (widening the spectral null width associated with an interferer).
In this paper, the focus is on high-dimensional setting where the dimension $p$ is high, while the data aspect ratio $n/p$ is low.
We propose an estimator called \tabasco{} (TApered or BAnded Shrinkage COvariance matrix) that  shrinks the tapered sample covariance matrix towards a scaled identity matrix.
We derive optimal and estimated (data adaptive) regularization parameters that are designed to minimize the mean squared error (MSE) between the proposed shrinkage estimator and the true covariance matrix.
These parameters are derived under the general assumption that the data is sampled from an unspecified elliptically symmetric distribution with finite 4th order moments (both real- and complex-valued cases are addressed). 
Simulation studies show that the proposed \tabasco{} outperforms all competing tapering covariance matrix estimators in diverse setups. 
A space-time adaptive processing (STAP) application also illustrates the benefit of the proposed estimator in a practical signal processing setup. 
\end{abstract}


% \fi
\IEEEpeerreviewmaketitle

%%%%%%%%%%%%%%%%%%%%%%%%%%%%%%%%%%%%%%%%%%%%%%%%%%%%%%%%%%%%%%%%%%%%%%%%%%%%%%%

%%%%%%%%%
% SECTION 1
%%%%%%%%%
\section{Introduction} \label{sec:intro}

Consider a set of $p$-dimensional (real-valued) vectors $\{ \x_i \}_{i=1}^n$ sampled from a distribution of a random vector $\x$ with unknown mean vector $\bom \mu=\E[\x]$ and unknown positive definite symmetric $p \times p$ covariance matrix $\M \equiv \mbox{cov}(\x)=\E[(\x-\bom \mu)(\x - \bom \mu)^\top]$. In the high-dimensional case and when the sample size $n$ is of the same order as $p$ ($p = \mathcal O(n)$) or $p \gg n$, one is required to use regularization (shrinkage) in order to improve the estimation accuracy of the SCM and to obtain a  positive definite matrix estimate. 
A popular estimate of $\M$ in such a setting is the regularized sample covariance matrix (RSCM), defined by 
\beq \label{eq:LW}
\S_\be = \be \S + (1-\be) \frac{\Tr(\S)}{p} \mathbf{I} , 
\eeq 
where $\be \in [0,1]$ is the regularization (or shrinkage) parameter, and where
\beq \label{eq:Mest}
\bo S = \frac{1}{n-1} \sum_{i=1}^n ( \x_i - \bar \x)(\x_i - \bar \x)^\top , 
\eeq 
denotes the unbiased sample covariance matrix (SCM), i.e., $\E[\S] = \M$. Note also that in \eqref{eq:Mest}, $\bar \x= \frac 1 n \sum_{i=1}^n \x_i$ denotes the sample mean vector.  
Automatic data-adaptive computation of optimal (oracle) parameter $\be$ for which $\S_\be$ in \eqref{eq:LW} attains the minimum mean squared error (MMSE) in Frobenius norm has been an active area of research. See for example  \cite{du2010fully,ledoit2004well,chen2010shrinkage,ollila2019optimal} to name only a few.

In many applications, the estimation accuracy (or another performance criterion) can alternatively be improved by using a so-called \textit{tapered} SCM.
Such estimate is defined as $ \bo W \circ \bo{S}$, where $\circ$ denotes the Hadamard (or Schur) element-wise product, and where $\bo W$ is a \textit{tapering} matrix (also referred to as covariance matrix taper), i.e., a template that imposes some additional structure to the SCM.  Note that above $(\bo{W} \circ \bo{S})_{ij} = w_{ij} s_{ij}$ for $(\bo{W})_{ij} = w_{ij} $ and $ (\bo{S})_{ij} = s_{ij} $. 

Covariance matrix tapers have been used in many applications in diverse fields. 
A first main example in statistics is related to covariance matrices with a diagonally dominant structure (e.g., in autoregressive models). 
This means that the variables have a natural order in the sense that $|i -j |$ large implies that the correlation between the $i$th and the $j$th variables is close to zero. 
In this settings, popular estimation approaches are to use a banding-type tapering matrices such as thresholding \cite{bickel2008regularized,bickel2008covariance}: 
\beq \label{eq:Wband}
(\bo{W})_{ij} = \begin{cases} 
 1, & | i - j | < k \\ 
0, & | i - j | \geq k
\end{cases} 
\eeq 
for some integer $k \in [\![1,p]\!]$ (called the bandwidth parameter), or softer thresholding variants.
Notably, the strong theoretical merits of a linear decay of the form
\beq \label{eq:Wminmax}
(\bo{W})_{ij}  = \begin{cases} 
 1, & | i - j | \leq k/2 \\ 
 2-   2 \dfrac{|i  - j |}{k} , &  k/2 < | i - j | < k \\ 
0, & | i -  j | \geq k
\end{cases} 
\eeq   
were studied in \cite{cai2010optimal}. 
A second major example concerns the signal processing literature, in which tapering matrices have been developed in order to improve several spectral properties of adaptive beamformers, or to compensate subspace leakage and calibration issues \cite{guerci2002principal}.
Most notably, the tapering matrices of the form
\beq \label{eq:Wsinc}
(\bo{W})_{ij}  =  {\rm sinc}( (i-j) \Delta / \pi )) 
\eeq 
where $\Delta \in \mathbb{R}^+$, attracted interest as a null broadening technique for fluctuating interference \cite{mailloux1995covariance,zatman1995production,guerci1999theory,song2003null,rugini2003regularized}.

A first approach to combine regularization with tapering was proposed in \cite{chen2012shrinkage} with the \textit{shrinkage to tapering} (ST) estimator, defined as the convex combination of the SCM and the tapered SCM: 
\beq  \label{eq:STest}
\S_{\text{ST},\be}  = \be \S + (1-\be)   (\W \circ \bo{S}), 
\eeq 
where $\be \in [0,1]$ is a shrinkage parameter. The authors then derived an optimal oracle parameter $\beta_o$ minimizing the MSE $\E[ \| \S_{\text{ST},\be}  - \M \|_{\Fr}^2]$ and proposed a shrinkage to tapering oracle approximating (STOA) estimator $\hat \beta_o$ of $\beta_o$ under the assumption of Gaussian data. 
Authors in  \cite{li2018estimation}  also studied the ST  estimator and derived 
an alternative oracle estimator of the shrinkage parameter both under Gaussian and non-Gaussian data.
Data adaptive selection of the bandwidth $k$ in \eqref{eq:Wband} was also addressed with cross validation \cite{chen2012shrinkage} or oracle estimation \cite{li2018estimation}.
A possible issue with the ST estimate is that it inherently destroys the tapering template structure (e.g., sparsity for banded matrices) since it can be expressed as  the modified tapered SCM $\S_{\text{ST},\be}  = ( \be \bo 1 \bo 1^\top + (1-\be)   \W  )\circ \bo{S} $. Hence, shrinkage is applied to the tapering matrix itself rather than to the SCM.
In the high dimensional case, it should also be noted that both $\bo W \circ \bo{S}$ and $\S_{\text{ST},\be}$ are not necessarily positive semidefinite matrices, i.e., they can have negative or null eigenvalues. 
A possible solution for this problem is to compute their EVD and then replacing the invalid eigenvalues by small positive constants. 
However, such a post-processing step further deteriorates the template pattern of the covariance matrix estimator, and is computationally restrictive when dealing with high-dimensional data.

In this paper we provide a solution to the aforementioned problems by jointly leveraging shrinkage to identity and tapering:
Let $\mathbb{W} = \{ \bo W (k) \}_{k=1}^K$ be a finite set of possible tapering matrices\footnote{
In this paper, we mostly focus on $k$ implying a notion of bandwidth (or model order), for which $\mathbb{W}$ can be constructed from \eqref{eq:Wband} or \eqref{eq:Wminmax} with $k\in[\![1,p ]\!]$.
However, the proposed methodology applies to the general setting where $\mathbb{W}$ corresponds to any finite collection of possibly envisioned templates.
Notably, we will also consider an application where $k$ indexes a set of possible 
$\{ \Delta_k \}_{k=1}^K$ used for the template model in \eqref{eq:Wsinc}.
}
satisfying $\W (k) \in  \mathcal W^+ ~ \forall k\in [\![1,K ]\!]$, with 
\beq
\label{eq:Set_W}
 \mathcal W^+ = \{ \W \in \Sym{p} : w_{ii}=1,  w_{ij} \geq 0  \,  \forall i, j  \in [\![1,p]\!]\}
\eeq
and with $ \Sym{p}$ denoting the set of all symmetric $p \times p$ matrices and $[\![1,p]\!] = \{1,\ldots, p\}$.
We propose an estimator, referred to as \tabasco{} (TApered or BAnded Shrinkage COvariance matrix), defined as
\beq
\hat \M_{\be,k}  =  \be (\bo W (k) \circ \bo{S} )  + (1-\be) \frac{\Tr(\S)}{p} \mathbf{I} , \label{eq:TAPSHRINK}
\eeq
which benefits both from shrinkage (as the classic estimator in \eqref{eq:LW}) and exploitation of structure via tapering.
Note that it also preserves the original scale of the SCM since $\Tr( \W \circ \S)= \Tr(\S)$ $\forall \W \in \mathcal W^+$. 
Obviously, the success of banding and/or tapering depends on one's ability to choose the parameters $\beta$ and $k$ correctly. 
In this scope, we derive a fully automatic data-adaptive evaluation of the optimal parameters that jointly  minimize the mean squared error $\E[ \| \hat \M_{\beta,k}  - \M \|_{\Fr}^2]$ under the general assumption that the data is sampled from an unspecified elliptically symmetric (ES) distribution with finite 4th order moments.
A main interest to consider the general ES model is that it encompasses the standard Gaussian one while still accounting for possibly heavy-tailed distribution. 
Thus this assumption yields robustness to a large class of possible underlying data distributions.
Our empirical experiments evidence that the proposed approach offers a near-to-optimal regularization parameter selection which outperform cross-validation schemes (especially at low sample support).
Since both the RSCM in \eqref{eq:LW} (if $\W= \bo 1 \bo 1^\top \in \mathbb{W}$) and the tapered SCM ($\beta=1$) appear as special cases of \eqref{eq:TAPSHRINK}, \tabasco{} performs never worse than these two estimators in terms of MSE independent of the underlying structure of the true covariance matrix.

The paper is structured as follows. 
In \autoref{sec:oracle} expressions for the oracle regularization parameters $\beta$ and $k$ that minimize the MSE are derived in the general case of sampling from an unspecified distribution with finite 4th-order moments.  
In \autoref{sec:tapMSE} we provide useful intermediate theoretical results about tapered SCM when the data is sampled from an unspecified ES distribution with finite 4th order moments. 
In \autoref{sec:pract_comp} a practical closed-form expression for the optimal regularization parameters are derived when sampling from an ES distribution, and an adaptive fully automatic procedure for their computation is proposed.
As it is shown that the optimal parameters depend on the sphericity of the tapered covariance matrix $\W \circ \M$, we addressed the estimation of this quantity in \autoref{sec:sphericity}. 
\autoref{sec:ext} extends our results to the special cases of known location ($\bmu=\bo 0$) and/or complex-valued observations.  
\autoref{sec:simul} provides simulation studies while in \autoref{sec:applic} the estimator is applied to STAP data. 
Finally,  \autoref{sec:concl}  concludes.  
The Appendix contains more technical proofs.

%%
%% SECTION 2
%%

\section{Oracle \tabasco{} parameters $\beta$ and $k$} \label{sec:oracle}

First, recall that the \tabasco{} estimator $\hat \M_{\be,k}$ is defined by \eqref{eq:TAPSHRINK} for a set $\mathbb{W} = \{ \bo W (k) \}_{k=1}^K$ of envisioned tapering matrices (cf. footnote $^1$ for examples) and a regularization parameter $\beta\in [0,1]$.
In this section, we derive the expression of the oracle parameters $\beta$ and $k$ that minimize MSE in the general case of sampling from an unspecified $p$-variate distribution with finite 4th-order moments.  

Before doing so, let us introduce some notations and statistical parameters that are elemental in the proposed method. 
The \emph{scale} and the \emph{sphericity} of $\M$ \cite{ledoit2002some,srivastava2005some} are denoted by 
\beq \label{eq:gamma} 
\eta= \frac{\tr(\M)}{p} \quad \mbox {and} \quad   \gamma \equiv \gamma(\M) =  \frac{p \tr(\M^2)}{\tr(\M)^2}  ,
\eeq 
respectively.
The scale corresponds to the mean of the eigenvalues of $\M$, while the sphericity measures how close $\M$ is to a scaled identity matrix: $\gamma \in [1,p]$,  where $\gamma=1$ if and only if $\M \propto \mathbf{I}$ and $\gamma = p$ if and only if $\M$ has its rank equal to 1. 
For any $\W \in \mathcal W^+$ as in \eqref{eq:Set_W},  the matrix $\W \circ \M$, is called the tapered covariance matrix and we denote 
\beq\label{eq:gamma_tap} 
\gamma_{\W} \equiv \gamma(\W \circ \M) =   \frac{p \tr\left((\W \circ \M)^2\right)}{\tr(\M)^2}, 
\eeq 
the sphericity parameter of the tapered covariance matrix.  
When $\W= \bo 1 \bo 1^\top$, we write $\gamma_{\bo 1 \bo 1^\top} \equiv \gamma$ for brevity. 

\subsection{Oracle shrinkage parameter $\beta$ for fixed $k$} 

We start by assuming that the index $k$ is fixed.
This allows us to simply denote the fixed tapering matrix $\W\equiv \W(k)$ and \tabasco{} as $\hat \M_\be\equiv \hat \M_{\be,k}$.
To find the oracle MMSE shrinkage parameter $\beta  \in [0,1]$ of $\hat \M_{\be}$, the aim is thus to solve  
\beq \label{eq:optimal_beta_o}
\beta_o =  \underset{\be\in [0,1]}{\arg \min}  \Big\{ \mathbb{E} \Big[ \big\| \hat{\M}_{\be} - \M \|^2_{\mathrm{F}} \Big] \Big\}, 
\eeq 
where $\| \cdot \|_{\Fr}$ denotes the \emph{Frobenius matrix norm}, i.e., $\| \mathbf{A} \|_{\Fr}^2 = \tr(\bo A^\top \bo A)$ 
and $\tr(\cdot)$ denotes the matrix trace, i.e., $\tr(\A)=\sum_{i} a_{ii}$ for all square matrices $\A=(a_{ij})$. 
 
Notice that the  MSE of the tapered SCM is 
\begin{align}
\mathrm{MSE}&(\W \circ \S) = \E \big[ \| \W \circ \S - \M \|_{\Fr}^2  \big]  \notag   \\ 
&= \E \left[  \right\| \W \circ \S \left \|^2_{\Fr} \right] + \| \M \|^2_{\Fr} - 2 \| \mathbf{V} \circ \M \|^2_{\Fr},  \label{eq:MSEtaper} 
\end{align}
where 
\beq \label{eq:VsqrtW}
\mathbf{V} = (v_{ij})_{p \times p}  \text{~with~} v_{ij} = \sqrt{w_{ij}} \text{~for~} \W \in \mathcal W^+.
\eeq 
By normalized MSE (NMSE) we refer to $\mathrm{NMSE}(\W \circ \S) = \MSE(\W \circ \S) / \| \M \|_{\Fr}^2$.  
We are now ready to state the main result of this section. 

\begin{theorem}  \label{th:beta0}    
Let $\{\x_i \}_{i=1}^n$ be an i.i.d. random sample from any $p$-variate distribution with finite 4th order moments. 
For any  fixed $\W \in \mathcal W^+$, the oracle  parameter $\be_o$ in \eqref{eq:optimal_beta_o}  is
\begin{align}
\be_o  &= \frac{ \left\| \bo V \circ\M - \eta \mathbf{I} \right\|_{\Fr}^2}{\E \Big [ \big\| \W \circ \S -   \eta \mathbf{I} \big\|_{\Fr}^2 \Big]} \label{eq:beta0id} \\ 
	    &= \frac{p (\gamma_{\V}-1) \eta^2}{\E \left[  \right\| \W \circ \S \left \|^2_{\Fr} \right] - p^{-1} \E[ \tr(\bo S)^2]} 
	    \label{eq:beta0id1}  \\
	    &= \frac{( \gamma_{\V}-1)}{ \gamma \cdot \mathrm{NMSE}( \W \circ \S) +   2 \gamma_{\V} -  \gamma -   \E[ \hat \eta^2] / \eta^2} \label{eq:beta0id2} 
\end{align}
where $\mathbf{V} = (v_{ij})$ with $v_{ij} = \sqrt{w_{ij}}$,  $\gamma_{\V}$ is defined via  \eqref{eq:gamma_tap} 
and $\hat \eta = \tr(\S)/p$. 
Furthermore, the value of the MSE
at the optimum  is
\begin{align} \label{eq:MSEopt}
 &\MSE(\hat{\M}_{\be_o})  =  \frac{\E\big[\tr (\S)^2  \big]  - \tr(\M)^2}{p}   \notag \\  
 &+ \| \M \|_{\Fr}^2 - \| \bo V \circ \M \|_{\Fr}^2  +  (1-\beta_0)\left\| \bo V \circ\M - \eta \mathbf{I} \right\|_{\Fr}^2.
\end{align}
\end{theorem}

\begin{proof}
The proof is postponed to Appendix~\ref{app:th:beta0}.
\end{proof}

Notice that \autoref{th:beta0}  also provides the optimal MMSE shrinkage parameter $\beta_o$ for the RSCM $\S_\be$ in  \eqref{eq:LW} 
since $\hat \M_{\be} = \S_{\be}$ when $\W= \bo 1 \bo 1^\top$.
For the RSCM the optimal parameter is 
\beq \label{eq:beta0_RSCM}
\beta_o= \frac{( \gamma-1)}{ \gamma \cdot \mathrm{NMSE}( \S) +    \gamma  -   \E[ \hat \eta^2]/\eta^2} , 
\eeq
where we used \eqref{eq:beta0id2}  and the facts that  $\gamma = \gamma_{\V}$ and $\W \circ \S = \S$ for  $\W= \bo 1 \bo 1^\top$.  The minimum MSE of the RSCM utilizing the optimal shrinkage parameter in  \eqref{eq:beta0_RSCM} is 
\begin{align*} 
\MSE(\S_{\be_o})  =  \frac{\E\big[\tr (\S)^2  \big]  - \tr(\M)^2}{p}   +  (1-\beta_0)\left\| \M - \eta \mathbf{I} \right\|_{\Fr}^2,
\end{align*}
where we used \eqref{eq:MSEopt} and that $\V \circ \M = \M$ for  $\V= \bo 1 \bo 1^\top$. 

\subsection{Oracle index $k$} 

Notice that $\MSE(\hat \M_{\be_0}) $ in \eqref{eq:MSEopt} implicitly depends on $k$ through $\W \equiv \W (k)$ and $\V$ defined in \eqref{eq:VsqrtW}.
We further have the relation
\begin{align} 
\mathrm{NMSE}(\hat{\M}_{\be_o}) 
 &= C  - \frac{ \| \bo V \circ \M \|_{\Fr}^2}{\|\M\|_{\Fr}^2}  +  (1-\beta_0)\frac{ \left\| \bo V \circ\M - \eta \mathbf{I} \right\|_{\Fr}^2}{\|\M\|_{\Fr}^2}  \notag   \\
&= C - \frac{\gamma_{\V}}{\gamma}  + (1-\beta_0 ) \frac{\gamma_{\V} - 1}{\gamma}  \notag \\
&=C  - \frac{1}{\gamma} + \frac{\beta_0 (1-\gamma_{\V})}{\gamma},  \label{eq:risk} 
\end{align}  
where $C$ is a constant that is not dependent on $k$. 
Equation \eqref{eq:risk} then implies that minimizing the MSE with respect to $k$ is equivalent to set
\beq \label{eq:optim_bandwidth}
k_o = \arg \underset{k}{\min}  \, \beta_0(k) (1-\gamma_{\V}(k)), 
\eeq 
where $\beta_o(k)$ is given by any of the expressions in \eqref{eq:beta0id}-\eqref{eq:beta0id2}  and $\gamma_{\V}(k)$ is defined via \eqref{eq:gamma_tap}. 
Note that we have made explicit the dependence of $\beta_0$ and $\gamma_{\V}$ on $k$ in \eqref{eq:optim_bandwidth} for clarity of exposition.

Of course, the oracles $\beta_0$ and $k_0$ depend here on the true underlying data distribution and covariance matrix through various unknown quantities.
A practical implementation of \tabasco{} thus requires their adaptive evaluation.
Rather than resorting to potentially inaccurate cross-validation, we will consider the general case where the data is sampled from an unspecified ES distribution \cite{fang1990symmetric,ollila2012complex}.
In this setting, we show that the oracle parameters eventually depend on few parameters that can be accurately evaluated, even at low sample support.

%% 
%%  SECTION 3
%% 
\section{Tapered SCM under ES distributions}  \label{sec:tapMSE}

In this section we recall some definitions and key results concerning ES distribution  \cite{fang1990symmetric,ollila2012complex}.
We then and derive useful results (expectations and consistent estimates) related to functions of the tapered SCM $\W \circ \S$, which will be needed in later developments of oracle \tabasco{} parameters. 

\subsection{ES distributions}

The probability density function of  an elliptically distributed random vector, denoted by $\x \sim \mathcal E_\pdim(\bom \mu,\M,g)$, is given by 
\beq \label{eq:pdf_ES} 
	f(\x)
	= C_{\pdim,g} |\M|^{-1/2} g\big( (\x-\bom \mu)^\top \M^{-1} (\x-\bom \mu)\big),
\eeq 
where $\M$ denotes the positive  definite symmetric covariance matrix parameter, $\bom \mu$ is the mean vector,  $g: \left[0,\infty\right) \to \left[0,\infty\right)$ is the \emph{density generator}, which is a fixed function that is independent of $\x, \bom \mu$ and $\M$, and $C_{\pdim,g}$ is a normalizing constant ensuring that $f(\x)$ integrates to 1. 
Note that here we define $g$ such that "scatter matrix" parameter $\M$ coincides with the covariance matrix. This can always be  assumed (under assumption of finite 2nd order moments) without any loss of generality  \cite{fang1990symmetric,ollila2012complex}.    For example, the multivariate normal (MVN) distribution, denoted by $\mathcal N_p(\bom \mu,\M)$, is obtained when $g(t)=\exp(-t/2)$.   
The flexibility regarding the density generator $g$ allows for modeling a large class of distributions, including heavy-tailed ones such as   the multivariate $t$-distribution (MVT) with $\nu>2$  degrees of freedom (d.o.f.), denoted by $\x \sim t_{\nu}(\bom \mu, \M)$,  where $\nu>2$ needs to be assumed  for finite 2nd-order moments.

The elliptical kurtosis~\cite{muirhead1982aspects} parameter $\kappa$ is defined as 
\beq \label{eq:kappa} 
 \ka=  
\dfrac{\E [r^4]}{p(p+2)}  - 1 = \frac 1 3 \mathrm{kurt}(x_i) ,
 \eeq  
 where the expectation is over the distribution of the random variable $r=  \| \M^{-1/2}(\x-\bom \mu) \|$ and $\mathrm{kurt}(x_i)$ denotes the excess kurtosis of any (e.g., $i$th) marginal variable of $\x$.    Furthermore, observe that $\E[r^2]=p$. 
 The elliptical kurtosis parameter vanishes (so $\ka=0$) when $\x$ has a MVN distribution. 

We also recall from \cite[Lemma~2]{ollila2019optimal} that 
\begin{align}
\E \left[ \left \| \S \right\|_{\Fr}^2 \right]  &= \left( 1 + \tau_1 + \tau_2 \right) \| \M \|_{\Fr}^2 +  \tau_1 \tr(\M)^2   ,  \label{eq:trS2} \\ 
\E \left [\tr(\bo S)^2 \right] 	&= 2 \tau_1  \| \M \|_{\Fr}^2 +   \big(1+ \tau_2 \big) \tr(\M)^2 \label{eq:trS_2} , 
\end{align}
where the scalars 
\beq \label{eq:tau_1and2}
\tau_1 = \frac{1}{n-1} + \frac{\kappa}{n} \qquad  \mbox{and} \quad  \tau_2 =  \frac{\kappa}{n}
\eeq 
are dependent on the elliptical distribution (and hence on the density generator $g$) only via its kurtosis parameter.

\subsection{Useful intermediate results about tapered SCM}

We now derive an extension of \cite[Lemma~2]{ollila2019optimal} for tapered SCM $\W \circ \S$.  
Write $\diag(\A)  \equiv \diag(a_{11}, \ldots, a_{pp})$ for any matrix $\A = (a_{ij})_{p\times p}$, where $\diag(\mathbf{a})$ denotes a diagonal matrix with the entries of vector $\a$ on the
main diagonal.
 
\begin{lemma} \label{lem:EtrTaper} Let $\{\x_i \}_{i=1}^n$ be an i.i.d. random sample from  $\mathcal E_p(\bmu, \M,g)$ with finite 4th order moments. Then for any $\W \in \mathcal W^+$, it holds that 
\[
\E \left[  \left\| \W \circ \S \right \|^2_{\Fr} \right]  =  (1+ \tau_1 + \tau_2) \| \W \circ \M \|^2_{\Fr}  +  \tau_1 \tr( (\D_{\M} \W)^2)
\]
and
\begin{align*}
\E &\left[  \tr( (\D_{\S} \W)^2 )\right]  = 2 \tau_1  \| \W \circ \M \|^2_{\Fr}  +  (1+\tau_2) \tr( (\D_{\M} \W)^2), 
\end{align*}
where $\D_{\M} = \diag(\M)$ and $\D_{\S} = \diag(\S)$. 
\end{lemma}

\begin{proof}
The proof is postponed to Appendix~\ref{app:lem:EtrTaper}.
\end{proof}

\noindent Note that if $\W= \bo 1 \bo 1^\top$, then $\tr( (\D_{\M} \W)^2 ) = \tr(\M)^2$ and $\W \circ \S  = \S$ so the expectations in \autoref{lem:EtrTaper} coincide with \cite[Lemma~2]{ollila2019optimal} (i.e., \eqref{eq:trS2} and \eqref{eq:trS_2}). 

Interestingly, the knowledge of $\E[  \left\| \W \circ \S \right \|^2_{\Fr}]$ from \autoref{lem:EtrTaper} allows for a direct computation of MSE of $ \W \circ \S$  via \eqref{eq:MSEtaper}. 
\autoref{lem:EtrTaper} also states that the obvious plug-in estimate  $\left\| \W \circ \S \right \|^2_{\Fr}/\pdim$ for the parameter 
\beq \label{eq:etatwo}
\etatwo =  \frac{\left\| \bo W \circ \M \right \|^2_{\Fr}}{p} 
\eeq 
is biased.  
Next we derive a proper estimator $\etatwohat$ of $\etatwo$ which  extends~\cite[Theorem~4]{ollila2019optimal} and provides an unbiased estimator of $\etatwo$ provided that the elliptical kurtosis parameter $\kappa$ is known.

\begin{theorem}\label{th:eta2ell} 
Let $\{\x_i \}_{i=1}^n$ be an i.i.d. random sample from a $p$-variate elliptical distribution $\mathcal E_p(\bom \mu, \M,g)$ with finite 4th order moments. Then, an unbiased
	estimator of $ \etatwo=  \| \bo W \circ \M \|^2_{\Fr}/p$ for any finite $n$ and $p$ and any $\W \in \mathcal W^+$ is 
  \begin{align*}
	 \etatwohat & =   b_n \left(\frac{\left\| \W \circ \S \right \|^2_{\Fr}}{p} -  a_n  \, \frac{\tr \left( (\D_{\S} \W)^2 \right)}{p} \right)   , 
  \end{align*}
  \vspace{-0.1cm}
where  
\begin{align} 
	a_n &=  \frac{1}{n+\kappa}  \left(  \frac{n}{n-1} + \kappa \right)  \label{eq:an}  \\ 
	b_n &= \frac{ (\ka  + n)(n-1)^2}{ (n-2)(3 \kappa (n-1) + n(n+1))}.  \label{eq:bn} 
\end{align}
\end{theorem}

\begin{proof}
Note that $a_n$ in \eqref{eq:an} can be written as $a_n = \tau_1/(1+\tau_2)$, where definitions of $\tau_1$ and $\tau_2$ are given by~\eqref{eq:tau_1and2} while $b_n$ in \eqref{eq:bn} can be expressed as $b_n = ( 1+\tau_1+\tau_2 - 2\tau_1 a_n \big)^{-1}$.
Then using \autoref{lem:EtrTaper},  we notice that 
\begin{align*}
&b_n^{-1} p \E[\etatwohat] = \left(\tau_1- a_n(1+\tau_2)\right) \tr( (\D_{\M} \W)^2 ) \\ 
	& \quad + \left(1+\tau_1+\tau_2-2\tau_1 a_n\right)  \left\| \W \circ \M \right \|^2_{\Fr} = b_n^{-1}  \left\| \W \circ \M \right \|^2_{\Fr} 
\end{align*} 
The expressions  \eqref{eq:an}  and  \eqref{eq:bn} are obtained when replacing the values of $\tau_1$ and $\tau_2$ given in \eqref{eq:tau_1and2} into $a_n\equiv a_n(\tau_1,\tau_2)$ and $b_n\equiv b_n(\tau_1,\tau_2)$ and simplifying the obtained expressions. 
\end{proof}

This result will notably be used later in \autoref{subsec:Ell2} to construct an estimator of the sphericity parameter $\gamma_{\W}$. 

%% 
%%  SECTION 4
%% 

\section{Oracle parameters estimation in ES distributions} \label{sec:pract_comp}

Using  \autoref{lem:EtrTaper} we may now derive a simple closed form expression of the optimal shrinkage parameter 
$\beta_o$ given in  \autoref{th:beta0}    
that depends only on  few summary (scalar-valued) statistics which can be estimated from the data.  
 Let us denote 
\beq \label{eq:thetaW}
\theta_{\W} =  \frac{\d^\top_{\M} (\W \circ \W) \d_{\M}}{p}  =  \frac{\tr( (\D_{\M} \W)^2 )}{p}, 
 \eeq 
where $\mathbf{d}_{\M} = (\sigma_1^2,\ldots,\sigma_p^2)^\top$ contains the variances of the variables, i.e., the diagonal elements of $\M$. 
The 2nd equality in \eqref{eq:thetaW} follows from \cite[Lemma~7.5.2]{horn2012matrix}. The main result of this section is derived next.

\begin{theorem}\label{th:beta0ell}
Let $\{\x_i \}_{i=1}^n$ be an i.i.d. random sample from an
ES distribution $\mathcal E_p(\bom \mu, \M,g)$ with finite 4th order moments.
For any $\W \in \mathcal W^+$, the oracle parameter $\be_o$ in \eqref{eq:optimal_beta_o} is
\begin{align} \label{eq:optim_beta0_ell}
\be_o   =    \dfrac{  t}{ t +   (n/(n-1))(\theta_\W/\eta^2  + \gamma_{\W} - 2 \gamma/p) + \kappa \cdot A   }    , 
\end{align}
 where  $t = n(\gamma_{\V}  -1)$, and 
\[
A=   \theta_{\W}/\eta^2 -1 + 2 \gamma_{\W}  - 2 \gamma/p.
\]
 \end{theorem}

\begin{proof} Follows from \autoref{th:beta0} after substituting the values of $\E \left[  \right\| \W \circ \S \left \|^2_{\Fr} \right] $ given in \autoref{lem:EtrTaper} and  of $\E\big[\tr(\S)^2\big]$  given  in  \eqref{eq:trS_2} into the denominator of $\be_o$ in \eqref{eq:beta0id1} and simplifying the expression.  
\end{proof}

Following from \autoref{th:beta0ell}, the proposed data-adaptive implementation of \tabasco{} consists in applying the oracle procedure of \autoref{sec:oracle} by replacing each of the unknown parameters $\{\eta,  \theta_{\W}, \kappa,\gamma,  \gamma_{\W}, \gamma_{\V} \}$ in \eqref{eq:optim_beta0_ell} by carefully chosen estimates (detailed below). This yields estimate $\hat \beta_o(k)$ and one considers 
all templates in set $\mathbb{W} = \{ \W(k) \}_{k=1}^K$.  
Similarly, the index $k$ is estimated based on \eqref{eq:optim_bandwidth} by replacing the unknown $\beta_0(k)$ and $\gamma_{\V}(k)$ by their estimates and solving
\beq \label{eq:hatk} 
\hat k_o  = \underset{ k }{ \arg \min}  \,  \hat \beta_0(k) (1- \hat \gamma_{\V}(k)). 
\eeq 
The pseudocode of the proposed estimation algorithm is summarized in \autoref{alg:tabasco}.

Estimators of the parameters $\{\eta,  \theta_{\W}, \kappa,\gamma,  \gamma_{\W}, \gamma_{\V} \}$ and additional remarks are detailed in the following:\\
$\bullet$ For $\eta$ and $\theta_{\W}$, we use the empirical estimates:
\beq \label{eq:eta_theta_hat}
\hat \eta= \tr(\S)/p \text{~and~}\hat \theta_{\W} =  \tr( (\D_{\S} \W)^2 )/p
\eeq
$\bullet$ The elliptical kurtosis $\kappa$ can be estimated using $\hat \kappa$ detailed in \cite[Sect.~IV]{ollila2019optimal} as (bias-corrected) average sample kurtosis of the marginal variables scaled by $1/3$. Also note that if the data is assumed to follow the MVN distribution, we can set $\kappa=0$, and the last term $\kappa \cdot A$ can be ignored in the denominator.\\
$\bullet$ The estimation of the three sphericity statistics: $\gamma$, $\gamma_{\W}$, and $\gamma_{\V}$ is addressed in detail in \autoref{sec:sphericity}.  
Also notice that $\V=(\sqrt{w_{ij}})_{p\times p}$, so if $\W$ is a selection matrix (i.e., that 
has only 0-s or 1-s as its off-diagonal elements), as for example in \eqref{eq:Wband}, then $\W = \V$ so only $\gamma_{\W}$ requires to be estimated.
 
\begin{algorithm}[!t]
\caption{ \tabasco{} }\label{alg:tabasco}
\DontPrintSemicolon
\SetKwInOut{Input}{Input}
\SetKwInOut{Output}{Output}
\SetKwInOut{Init}{Initialize}
\SetNlSkip{0.3em}
\SetInd{0.5em}{0.5em}
\Input{Data $\{\x_i \}_{i=1}^n$, templates set $\{ \W (k) \}_{k=1}^K$}

Compute SCM $\bf S$ in \eqref{eq:Mest} and SSCM $\hat \La$ in \eqref{eq:SSGN} 

Compute $\hat \eta$ from \eqref{eq:eta_theta_hat}

Compute $\hat \kappa$ from \cite[Sect.~IV]{ollila2019optimal}

Compute $\hat \gamma$ (options Ell1- or Ell2- in \autoref{sec:sphericity})

\BlankLine
 
\For{ $k\in [\![1,K ]\!]$ }{

Set $\W = \W(k)$ and $\V = \V(k)=(\sqrt{w_{ij}(k)})_{p \times p }$ 

Compute $\hat \theta_{\W} $ from \eqref{eq:eta_theta_hat}

Compute $\hat \gamma_{\W}(k)$ and $ \hat \gamma_{\V}(k)$ (options in \autoref{sec:sphericity})

Compute  $\hat \be_o(k) $ from \eqref{eq:optim_beta0_ell} using plug-in estimates

} 

Select optimal $k_0$ as in \eqref{eq:hatk} with $\{ \hat \be_o(k) , \hat \gamma_{\V}(k) \}_{k=1}^K$

Set $\W = \W( \hat k_o)$ and $\hat \beta = \hat \beta_o(k)$

\Output{$\hat \M  =  \hat \be \cdot  (\bo W \circ \bo{S} )  + (1- \hat \be) \hat \eta \mathbf{I}$} 

\end{algorithm}

%% 
%%  SECTION 5
%% 
\section{Estimators of sphericity} \label{sec:sphericity} 

In this section, we detail two new alternative estimators of the sphericity of the tapered covariance matrix $\W \circ \M$, which are extensions of the sphericity estimators proposed in \cite{ollila2019optimal}.  First,  define the shape matrix (or normalized covariance matrix) as
$  \shape = p \frac{\M}{\tr(\M)}$ and note that $\tr(\shape)=p$. The sphericity measures 
$\gamma$ and $\gamma_{\W}$ for any $\W \in \mathcal W^+$ 
can then be expressed simply in terms of $\shape$ via the formulas: 
\[
\gamma =  \frac{ \| \shape\|_{\Fr}^2}{p} \quad \mbox{and} \quad  \gamma_{\W} = \frac{\| \W \circ \shape \|_{\Fr}^2}{p}.
\]

\subsection{Ell1-estimator of sphericity}

The Ell1-estimator is based on the \emph{spatial sign covariance matrix} (SSCM), which has been popular for constructing robust estimates of the sphericity \cite{zou2014multivariate, zhang2016automatic}.
This estimator was theoretically studied in \cite{raninen2020linear} and we propose here its adaptation to the sphericity of the tapered covariance matrix $\W \circ \M$.

The (scaled) SSCM is defined by 
\beq \label{eq:SSGN} 
\hat \La = \frac{p}{n} \sum_{i=1}^{n}  \frac{ (\x_{i} - \hat{\bom \mu})(\x_{i} - \hat{\bom \mu})^\top}{\|\x_{i} - \hat{\bom \mu} \|^{2}},  
\eeq 
where $\hat{\bom{\mu}} = \arg \min_{\bom \mu} \sum_{i=1}^{n} \| \x_{i} - \bom \mu \|$ is
the \emph{sample spatial median}~\cite{brown1983statistical}.
When $\bmu$ is known (and without loss of generality assuming $\bmu = \bo 0$), the SSCM is defined as 
$
\hat \La = \frac{p}{n} \sum_{i=1}^{n} \frac{ \x_{i}\x_{i}^\top}{\|\x_{i} \|^{2}}.  
$
Recently, it was shown  in \cite{raninen2020linear} that the following estimate of sphericity based on the SSCM (when $\bom{\mu}$ is known), 
\begin{equation}\label{eq:gammahat}
    \hat \gamma =  \frac{n}{n-1}\left(   \frac{\| \hat \La \|_{\Fr}^2}{p}  -    \frac{p}{n}\right), 
\end{equation}
is asymptotically (as $p \to \infty$) unbiased when sampling from elliptical distributions under the following assumption 
\begin{itemize}
\item[(A)] The sequence of covariance matrix structures being considered
with increasing $p$ satisfies  $\gamma =  o(p)$ as $p\to \infty$.
\end{itemize}
In other words,  $\E[ \hat \gamma ]  \to \gamma $ as $p \to \infty$ when (A) holds.  
We note that Assumption~(A) is sufficiently general and holds for many covariance matrix models as shown in \cite[Prop.~3]{raninen2020linear}.  
The following Theorem presents a modification of the Ell1-estimator  \cite{ollila2019optimal} for the sphericity of $\W \circ \M$ with equivalent asymptotic guarantees.

\begin{theorem} \label{th:Ell1}
Let $\{\x_i \}_{i=1}^n$ be an i.i.d. random sample from an ES distribution $\mathcal E_p(\bom \mu, \M,g)$ with known $\bmu = \bo 0$.  
Then, for any $\W \in \mathcal W^+$ and under Assumption~(A),  the following statistic 
 \begin{align}\label{eq:gammahat_tap}
\hat \gamma_{\W} =   \frac{n}{n-1}\left(  \frac{ \| \bo W \circ  \hat \La \|_{\Fr}^2}{p}  - \frac{ \tr \left( ( \bo D_{\hat \La} \mathbf{W})^2 \right)}{np} \right ) , 
\end{align}
where $\bo D_{\hat \La}  = \diag(\hat \La)$,  is asymptotically,  as $p \to \infty$, unbiased estimator  of $ \gamma_{\W} = \gamma(\W \circ \M)$ in 
\eqref{eq:gamma_tap}, i.e.,   $\E[ \hat \gamma_{\W} ]  \to \gamma_{\W} $ as $p \to \infty$, for any fixed $n$. 
\end{theorem}

\begin{proof} Proof is postponed to the Appendix~\ref{app:th:Ell1}.
\end{proof} 

Observe that when $\bo W = \bo 1  \bo 1^\top$, then  $\hat \gamma_{\W}$ reduces to $\hat \gamma$ in  \eqref{eq:gammahat}. 

\subsection{Ell2-estimator of sphericity} \label{subsec:Ell2}

The Ell2-estimator of sphericity was proposed in \cite{ollila2019optimal} and we derive here its adaptation to the sphericity of the tapered covariance matrix $\W \circ \M$ thanks to \autoref{th:eta2ell}.

First, note that the sphericity of tapered covariance matrix can also be written as 
\[
\gamma_{\W}  = \etatwo/\eta^2, 
\]
where $\etatwo$ and  $\eta$ are defined in \eqref{eq:etatwo} and \eqref{eq:gamma} respectively. 
Using this expression, we consider the estimate where $\etatwohat$ is computed from \autoref{th:eta2ell}, and $\hat \eta^2$ is obtained from \eqref{eq:eta_theta_hat}.
This yields the estimator 
\begin{equation}\label{eq:hatgamma_ell2}
	\hat \gamma_{\W}
	=   p \hat b_n \left( \frac{\left\| \W \circ \S \right \|^2_{\Fr}}{ \tr(\S)^2}  -  \hat a_n  \frac{\tr\left( (\D_{\S} \W)^2 \right) }{\tr(\S)^2} \right), 
\end{equation}
where $\hat a_n \equiv a_n(\hat \kappa)$ and $\hat b_n \equiv b_n (\hat \kappa)$ are obtained by replacing the unknown $\kappa$ in \eqref{eq:an} and \eqref{eq:bn} by its estimate $\hat \kappa$ \cite[Sect.~IV]{ollila2019optimal}.  
We refer to \eqref{eq:hatgamma_ell2} as Ell2-estimator of sphericity $\gamma_{\W}$.
Also note that, if $n$ is  reasonably large, then  $ \hat b_n \approx 1$ and $n/( n+ \hat \kappa) \approx 1$,  its expression can be simplified to
\[
\hat \gamma_{\W}  \approx \frac{ p \left\| \W \circ \S \right \|^2_{\Fr}}{ \tr(\S)^2}  -
  (1+ \hat \kappa) \frac{p}{n}   \frac{\tr( (\D_{\S} \W)^2) }{\tr(\S)^2} . 
 \]
In the non-tapered case ($\W= \bo 1 \bo 1^\top$), the estimator in \eqref{eq:hatgamma_ell2} reduces to the Ell2-estimator of sphericity in \cite{ollila2019optimal}.
  
Although Ell2-estimator of sphericity does not require knowledge of the underlying elliptically symmetric distribution of the data, it is not a robust estimator.  
Thus we overall favour Ell1-estimator due to robustness of SSCM, and recommend usage of Ell2-estimator when dealing with data that is not heavy-tailed, i.e., which can be approximated by a Gaussian  distribution. 
In practice, we also always use the thresholding
\begin{equation}\label{eq:hatgamma_ell1_eq2}
   \hat \gamma
   = \min(p, \max(1, \hat \gamma))
\end{equation}
for any option in order to guarantee that the final estimator remain in the valid interval, $1\leq \gamma \leq p$.
  
%% 
%%  SECTION 6
%% 
\section{Extensions and special cases}  \label{sec:ext}

\subsection{Known location $\bom \mu $} 

In some applications, the mean vector $\bom \mu = \E[\x]$ is known and assumed to be $\bom \mu = \bo 0$ without loss of generality.
In this case, the covariance matrix $\M= \E[\x\x^\top]$ is estimated by the SCM,  defined by 
\beq \label{eq:S_known_location} 
\bo S  = \frac{1}{n} \sum_{i=1}^n  \x_i \x_i^\top. 
\eeq 
which is also unbiased estimator of $\M$, i.e., $\E[\S]=\M$. 
The known location case implies only small changes in our estimation procedure since \autoref{th:beta0}  holds for both known and unknown location cases.   

When the location is known, the expectation $\E \big[ \| \S \|_{\Fr}^2 \big] $ and $\E \left [\tr(\bo S)^2 \right]$  are of the form  \eqref{eq:trS2} and \eqref{eq:trS_2} with $\tau_1$ and $\tau_2$ given by 
 \beq \label{eq:tau_1and2_muknown}
\tau_1 = \frac{1+\kappa}{n} \qquad  \mbox{and} \quad  \tau_2 =  \frac{\kappa}{n}.
\eeq 
This result follows as a special case of \cite[Lemma~1]{ollila2021shrinking} for a Gaussian  weight function.    Similarly~\autoref{lem:EtrTaper} holds when using $\tau_1$ and $\tau_2$ in \eqref{eq:tau_1and2_muknown}.  The change to the optimal $\beta_0$ parameter is also minimal: 
one may ignore the term $(n/(n-1))$  that appears as the multiplier of the 2nd last term $\theta_\W/\eta^2 + \gamma_{\W} - 2 \gamma/p$ in the denominator of $\beta_0$. 
Theorem~\ref{th:eta2ell} also holds with  
\begin{align*} \label{eq:bn_known_location} 
a_n =  \frac{1  + \kappa}{n+\kappa } \qquad \mbox{ and } \quad b_n = \frac{  n (n + \ka)}{ (n-1)(n+2+3 \kappa)} .
\end{align*} 
% The {\tt tabasco} function in our Matlab Toolbox (\url{https://xxx}) allows the user to specify if the location  $\bmu$ is known in the model. 
 
 \subsection{Complex-valued data} 
  
Extending the results to complex-valued data also requires minor adaptations since \autoref{th:beta0} holds for complex-valued observations as well.  
First we recall some notations specific to complex-valued case. 
By $\| \x \|^2 = \x^{\hop} \x$ we denote the usual Euclidean norm in complex vector spaces, while $\| \bo B \|_{\Fr} =  \sqrt{\Tr(\bo B^\hop \bo B)}$ denotes the  Frobenius norm of a matrix $\bo B \in \C^{m \times n}$, where  
 $(\cdot)^\hop = [( \cdot)^*]^\top$ denotes the conjugate transpose  (or Hermitian transpose).  For any $x \in \C$, the notation  $| \cdot |$ refers to modulus, so $| x |^2 = x  x^*$

We now assume that the data $\{\x_i \}_{i=1}^n$ is  a random sample from a  circular complex elliptically symmetric (CES) distribution, denoted $\x \sim \C \mathcal E_p(\bmu, \M, g)$ (cf. \cite{ollila2012complex} for a detailed review).
Similarly to the real-valued case, the probability density function of a CES distributed random vector  $\x \in \C^p$ is given by
\begin{equation*}
f( \x ) =
C_{p,g}    |\M|^{-1} g((\x - \bmu)^\hop \M^{-1} (\x - \bmu)),
\end{equation*}
where $\M$ denotes the positive definite Hermitian covariance matrix, $\bmu = \E[\x]$ is the mean vector, $g: \R_{\geq 0} \to \R_{>0}$ is the density generator, and $C_{p,g} $ is a normalizing constant.
Again, we also normalize $g$ so that $\M = \E[ (\x - \bmu)(\x - \bmu)^\hop]$. 
The definitions of the scale and sphericity parameters in \eqref{eq:gamma} and \eqref{eq:gamma_tap} remain unchanged.  
The elliptical kurtosis is however re-defined as 
\begin{equation*}
    \kappa =  \dfrac{\E [r^4]}{p(p+1)}  - 1 = \frac 1 2 \text{kurt}(x_i).
\end{equation*}
where the expectation is over $r=  \| \M^{-1/2}(\x-\bom \mu) \|$ and $\mathrm{kurt}(x_i)$ denotes the excess kurtosis of any (e.g., $i$th) marginal variable of $\x$, defined by 
\begin{equation*}
    \text{kurt}(x_i) = \frac{\E[|x_i-\mu_i|^4]}{ \sigma_i^4} - 2,
\end{equation*}
where $\mu_i = \E[x_i]$ and $\sigma_i^2 = \E[|x_i-\mu_i|^2]$ denote the mean and variance of $x_i$. 
The theoretical lower bound of the kurtosis in the complex-valued case is $\kappa^{\text{LB}} =
-1/(p+1)$~\cite{ollila2012complex}.     Again $\kappa=0$ if $\x$ has a circular complex multivariate normal distribution ($\x \sim \mathbb{C} \mathcal N_p(\bmu,\M)$).
The SCM~\eqref{eq:Mest} of complex-valued observations is defined by 
\beq \label{eq:Mest_complex}
\bo S  = \frac{1}{n-1} \sum_{i=1}^n ( \x_i - \bar \x)(\x_i - \bar \x)^\hop 
\eeq 
and the \tabasco{} estimator $\hat \M_{\be}$ is still defined as in \eqref{eq:TAPSHRINK}. 
The next result provides the complex-valued extension of \autoref{lem:EtrTaper}. 
  
\begin{lemma} \label{lem:EtrTaper_c} Let $\{\x_i \}_{i=1}^n$ be an i.i.d.\ random sample from  $ \C \mathcal E_p(\bmu, \M,g)$ with finite 4th order moments. Then for any $\W \in \mathcal W^+$, 
and for the SCM as in \eqref{eq:Mest_complex}, it holds that 
\[
\E \left[  \left\| \W \circ \S \right \|^2_{\Fr} \right]  
=  (1+  \tau_2) \| \W \circ \M \|^2_{\Fr}  +  \tau_1 \tr( (\D_{\M} \W)^2)
\]
and
\begin{align*}
\E &\left[  \tr( (\D_{\S} \W)^2 )\right]  
= 2 \tau_1  \| \W \circ \M \|^2_{\Fr}  +  (1+\tau_2) \tr( (\D_{\M} \W)^2), 
\end{align*}
where $\D_{\M} = \diag(\M)$, $\D_{\S} = \diag(\S)$ and $\tau_1$ and $\tau_2$ are defined in \eqref{eq:tau_1and2}.
\end{lemma}

\begin{proof}
The proof is postponed to Appendix~\ref{app:lem:EtrTaper_c}.
\end{proof}

This result allows us to derive the complex-valued counterpart of \autoref{th:beta0ell} for the optimal shrinkage parameter $\beta_o$. 

\begin{theorem}\label{th:beta0ell_c}
Let $\{\x_i \}_{i=1}^n$ be an i.i.d. random sample from a 
complex elliptical distribution $ \C \mathcal E_p(\bom \mu, \M,g)$ with finite 4th order moments.  Then the oracle parameter $\be_o$  in \eqref{eq:optimal_beta_o}  is
\begin{align*} 
 \be_o   &=    \dfrac{  t}{ t +   (n/(n-1))( \theta_\W/\eta^2   - 2 \gamma/p) + \kappa \cdot A   }    , 
\end{align*}
 where  $t = n(\gamma_{\V}  -1)$, and 
\[
A=   \theta_{\W}/\eta^2 -1 +  \gamma_{\W}  - 2 \gamma/p.
\]
 \end{theorem}

With similar arguments as in the real-valued case, it follows  that \autoref{th:eta2ell} holds with $a_n$ as in \eqref{eq:an}  and $b_n$ given by 
\[
b_n =  \frac{ n (n-1)^2  (\kappa + n)}{  2 \kappa n (n^2 - 4n + 3) - \kappa^2 (n-1)^2 + n^2 ( n^2 - 2n  - 1)}. 
\]
This means that Ell2-sphericity estimator can be defined as earlier with changes only in equations for $a_n$ and $b_n$. 
Similarly, the only change for SSCM in~\eqref{eq:SSGN} for  complex-valued observations is that the transpose $(\cdot)^\top$ is replaced with the Hermitian transpose. 

%% 
%%  SECTION 7
%% 

\section{Simulation studies}  \label{sec:simul}

We generate samples from (real-valued) ES distributions with a scatter  matrix  $\M$  having a diagonally dominant structure (model 1 and model 2 detailed below). 
The mean $\bom \mu$ is generated randomly as $\mathcal N_p(10 \cdot \bo 1, \bo I)$ and  the number of Monte-Carlo trials is 5000.

The estimators included in the study are:
$i$) The Ledoit-Wolf estimator ({\bf LWE}) \cite{ledoit2004well} defined by \eqref{eq:LW} where $\beta$ is an estimate of an (oracle) MMSE parameter $\beta_o$. 
$ii$) The shrinkage to tapering oracle approximate ({\bf STOA}) estimator \cite{chen2012shrinkage} 
defined by \eqref{eq:STest} where $\beta$ is an estimate of the  oracle parameter computed using an iterative procedure. The bandwidth $k$ is selected using a cross-validation scheme with  60\%-to-40\% split for training and testing. 
$iii$) The shrinkage to tapering (ST-)estimators in \cite{li2018estimation} defined by \eqref{eq:STest} where both $\beta$ and $k$ are estimates of the  oracle MMSE parameters. 
The estimator {\bf ST-gaus} assumes Gaussian data, while {\bf ST-nong} assumes non-Gaussian (ES) data. 
$iv$) \tabasco{} (computed via \autoref{alg:tabasco}) using the Ell1-estimator of sphericity. 

\subsection{Model 1}  

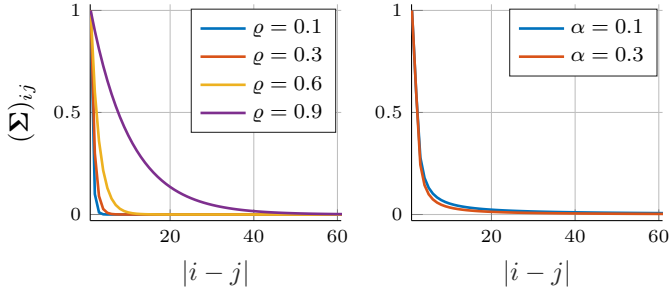
\begin{figure}[!t]
\centering
\setlength\fwidth{0.23\textwidth}
\centerline{
\definecolor{mycolor1}{rgb}{0.00000,0.44700,0.74100}%
\definecolor{mycolor2}{rgb}{0.85000,0.32500,0.09800}%
\definecolor{mycolor3}{rgb}{0.92900,0.69400,0.12500}%
\definecolor{mycolor4}{rgb}{0.49400,0.18400,0.55600}%
\begin{tikzpicture}

\begin{axis}[%
width=0.8\fwidth,
%height=0.33\fwidth,
at={(0\fwidth,0\fwidth)},
scale only axis,
xmin=0.9,
 tick label style={font=\scriptsize} , 
xmax=60.9,
xlabel style={font=\color{white!15!black}},
xlabel={$ |i - j|$},
ylabel={$(\M)_{ij}$},
ymin=-0.03,
ymax=1.03,
axis background/.style={fill=white},
axis x line*=bottom,
axis y line*=left,
xmajorgrids,
ymajorgrids,
legend style={legend cell align=left, align=left, draw=white!15!black,font = {\footnotesize}}
]
\addplot [color=mycolor1, line width=1.0pt]
  table[row sep=crcr]{%
1	1\\
2	0.1\\
3	0.01\\
4	0.001\\
5	0.0001\\
6	1e-05\\
7	1e-06\\
8	1e-07\\
9	1e-08\\
10	1e-09\\
11	1e-10\\
12	1e-11\\
13	1e-12\\
14	1e-13\\
15	1e-14\\
16	1e-15\\
17	1e-16\\
18	1e-17\\
19	1e-18\\
20	1e-19\\
21	1e-20\\
22	1e-21\\
23	1e-22\\
24	1e-23\\
25	1e-24\\
26	1e-25\\
27	1e-26\\
28	1e-27\\
29	1e-28\\
30	1e-29\\
31	1e-30\\
32	1e-31\\
33	1e-32\\
34	1e-33\\
35	1e-34\\
36	1e-35\\
37	1e-36\\
38	1e-37\\
39	1e-38\\
40	1e-39\\
41	1e-40\\
42	1e-41\\
43	1e-42\\
44	1e-43\\
45	1e-44\\
46	1e-45\\
47	1e-46\\
48	1e-47\\
49	1e-48\\
50	1e-49\\
51	1e-50\\
52	1e-51\\
53	1e-52\\
54	1e-53\\
55	1e-54\\
56	1e-55\\
57	1e-56\\
58	1e-57\\
59	1e-58\\
60	1e-59\\
61	1e-60\\
62	1e-61\\
63	1e-62\\
64	1e-63\\
65	1e-64\\
66	1e-65\\
67	1e-66\\
68	1e-67\\
69	1e-68\\
70	1e-69\\
71	1e-70\\
72	1e-71\\
73	1e-72\\
74	1e-73\\
75	1e-74\\
76	1e-75\\
77	1e-76\\
78	1e-77\\
79	1e-78\\
80	1e-79\\
81	1e-80\\
82	1e-81\\
83	1e-82\\
84	1e-83\\
85	1e-84\\
86	1e-85\\
87	1e-86\\
88	1e-87\\
89	1e-88\\
90	1e-89\\
91	1.00000000000001e-90\\
92	1.00000000000001e-91\\
93	1.00000000000001e-92\\
94	1.00000000000001e-93\\
95	1.00000000000001e-94\\
96	1.00000000000001e-95\\
97	1.00000000000001e-96\\
98	1.00000000000001e-97\\
99	1.00000000000001e-98\\
100	1.00000000000001e-99\\
};
\addlegendentry{$\varrho=0.1$}

\addplot [color=mycolor2, line width=1.0pt]
  table[row sep=crcr]{%
1	1\\
2	0.3\\
3	0.09\\
4	0.027\\
5	0.0081\\
6	0.00243\\
7	0.000729\\
8	0.0002187\\
9	6.561e-05\\
10	1.9683e-05\\
11	5.9049e-06\\
12	1.77147e-06\\
13	5.31441e-07\\
14	1.594323e-07\\
15	4.782969e-08\\
16	1.4348907e-08\\
17	4.3046721e-09\\
18	1.29140163e-09\\
19	3.87420489e-10\\
20	1.162261467e-10\\
21	3.486784401e-11\\
22	1.0460353203e-11\\
23	3.1381059609e-12\\
24	9.41431788269999e-13\\
25	2.82429536481e-13\\
26	8.47288609442999e-14\\
27	2.541865828329e-14\\
28	7.62559748498699e-15\\
29	2.2876792454961e-15\\
30	6.86303773648829e-16\\
31	2.05891132094649e-16\\
32	6.17673396283946e-17\\
33	1.85302018885184e-17\\
34	5.55906056655552e-18\\
35	1.66771816996665e-18\\
36	5.00315450989996e-19\\
37	1.50094635296999e-19\\
38	4.50283905890997e-20\\
39	1.35085171767299e-20\\
40	4.05255515301897e-21\\
41	1.21576654590569e-21\\
42	3.64729963771707e-22\\
43	1.09418989131512e-22\\
44	3.28256967394537e-23\\
45	9.8477090218361e-24\\
46	2.95431270655083e-24\\
47	8.86293811965249e-25\\
48	2.65888143589575e-25\\
49	7.97664430768724e-26\\
50	2.39299329230617e-26\\
51	7.17897987691851e-27\\
52	2.15369396307555e-27\\
53	6.46108188922666e-28\\
54	1.938324566768e-28\\
55	5.81497370030399e-29\\
56	1.7444921100912e-29\\
57	5.23347633027359e-30\\
58	1.57004289908208e-30\\
59	4.71012869724623e-31\\
60	1.41303860917387e-31\\
61	4.23911582752161e-32\\
62	1.27173474825648e-32\\
63	3.81520424476945e-33\\
64	1.14456127343083e-33\\
65	3.4336838202925e-34\\
66	1.03010514608775e-34\\
67	3.09031543826325e-35\\
68	9.27094631478976e-36\\
69	2.78128389443693e-36\\
70	8.34385168331078e-37\\
71	2.50315550499324e-37\\
72	7.50946651497971e-38\\
73	2.25283995449391e-38\\
74	6.75851986348173e-39\\
75	2.02755595904452e-39\\
76	6.08266787713356e-40\\
77	1.82480036314007e-40\\
78	5.4744010894202e-41\\
79	1.64232032682606e-41\\
80	4.92696098047818e-42\\
81	1.47808829414345e-42\\
82	4.43426488243036e-43\\
83	1.33027946472911e-43\\
84	3.99083839418733e-44\\
85	1.1972515182562e-44\\
86	3.59175455476859e-45\\
87	1.07752636643058e-45\\
88	3.23257909929173e-46\\
89	9.6977372978752e-47\\
90	2.90932118936256e-47\\
91	8.72796356808768e-48\\
92	2.6183890704263e-48\\
93	7.85516721127891e-49\\
94	2.35655016338367e-49\\
95	7.06965049015102e-50\\
96	2.12089514704531e-50\\
97	6.36268544113592e-51\\
98	1.90880563234078e-51\\
99	5.72641689702233e-52\\
100	1.7179250691067e-52\\
};
\addlegendentry{$\varrho=0.3$}

\addplot [color=mycolor3, line width=1.0pt]
  table[row sep=crcr]{%
1	1\\
2	0.6\\
3	0.36\\
4	0.216\\
5	0.1296\\
6	0.07776\\
7	0.046656\\
8	0.0279936\\
9	0.01679616\\
10	0.010077696\\
11	0.0060466176\\
12	0.00362797056\\
13	0.002176782336\\
14	0.0013060694016\\
15	0.00078364164096\\
16	0.000470184984576\\
17	0.0002821109907456\\
18	0.00016926659444736\\
19	0.000101559956668416\\
20	6.09359740010496e-05\\
21	3.65615844006297e-05\\
22	2.19369506403778e-05\\
23	1.31621703842267e-05\\
24	7.89730223053602e-06\\
25	4.73838133832161e-06\\
26	2.84302880299297e-06\\
27	1.70581728179578e-06\\
28	1.02349036907747e-06\\
29	6.14094221446481e-07\\
30	3.68456532867889e-07\\
31	2.21073919720733e-07\\
32	1.3264435183244e-07\\
33	7.95866110994639e-08\\
34	4.77519666596784e-08\\
35	2.8651179995807e-08\\
36	1.71907079974842e-08\\
37	1.03144247984905e-08\\
38	6.18865487909431e-09\\
39	3.71319292745659e-09\\
40	2.22791575647395e-09\\
41	1.33674945388437e-09\\
42	8.02049672330623e-10\\
43	4.81229803398374e-10\\
44	2.88737882039024e-10\\
45	1.73242729223415e-10\\
46	1.03945637534049e-10\\
47	6.23673825204292e-11\\
48	3.74204295122575e-11\\
49	2.24522577073545e-11\\
50	1.34713546244127e-11\\
51	8.08281277464763e-12\\
52	4.84968766478858e-12\\
53	2.90981259887314e-12\\
54	1.74588755932389e-12\\
55	1.04753253559433e-12\\
56	6.28519521356599e-13\\
57	3.7711171281396e-13\\
58	2.26267027688376e-13\\
59	1.35760216613025e-13\\
60	8.14561299678152e-14\\
61	4.88736779806892e-14\\
62	2.93242067884135e-14\\
63	1.75945240730481e-14\\
64	1.05567144438289e-14\\
65	6.33402866629731e-15\\
66	3.80041719977839e-15\\
67	2.28025031986703e-15\\
68	1.36815019192022e-15\\
69	8.20890115152132e-16\\
70	4.92534069091279e-16\\
71	2.95520441454767e-16\\
72	1.7731226487286e-16\\
73	1.06387358923716e-16\\
74	6.38324153542297e-17\\
75	3.82994492125378e-17\\
76	2.29796695275227e-17\\
77	1.37878017165136e-17\\
78	8.27268102990817e-18\\
79	4.9636086179449e-18\\
80	2.97816517076694e-18\\
81	1.78689910246017e-18\\
82	1.0721394614761e-18\\
83	6.4328367688566e-19\\
84	3.85970206131396e-19\\
85	2.31582123678837e-19\\
86	1.38949274207302e-19\\
87	8.33695645243815e-20\\
88	5.00217387146289e-20\\
89	3.00130432287773e-20\\
90	1.80078259372664e-20\\
91	1.08046955623598e-20\\
92	6.4828173374159e-21\\
93	3.88969040244954e-21\\
94	2.33381424146972e-21\\
95	1.40028854488183e-21\\
96	8.40173126929101e-22\\
97	5.0410387615746e-22\\
98	3.02462325694476e-22\\
99	1.81477395416686e-22\\
100	1.08886437250011e-22\\
};
\addlegendentry{$\varrho=0.6$}

\addplot [color=mycolor4, line width=1.0pt]
  table[row sep=crcr]{%
1	1\\
2	0.9\\
3	0.81\\
4	0.729\\
5	0.6561\\
6	0.59049\\
7	0.531441\\
8	0.4782969\\
9	0.43046721\\
10	0.387420489\\
11	0.3486784401\\
12	0.31381059609\\
13	0.282429536481\\
14	0.2541865828329\\
15	0.22876792454961\\
16	0.205891132094649\\
17	0.185302018885184\\
18	0.166771816996666\\
19	0.150094635296999\\
20	0.135085171767299\\
21	0.121576654590569\\
22	0.109418989131512\\
23	0.0984770902183612\\
24	0.0886293811965251\\
25	0.0797664430768726\\
26	0.0717897987691853\\
27	0.0646108188922668\\
28	0.0581497370030401\\
29	0.0523347633027361\\
30	0.0471012869724625\\
31	0.0423911582752162\\
32	0.0381520424476946\\
33	0.0343368382029252\\
34	0.0309031543826326\\
35	0.0278128389443694\\
36	0.0250315550499324\\
37	0.0225283995449392\\
38	0.0202755595904453\\
39	0.0182480036314007\\
40	0.0164232032682607\\
41	0.0147808829414346\\
42	0.0133027946472911\\
43	0.011972515182562\\
44	0.0107752636643058\\
45	0.00969773729787525\\
46	0.00872796356808772\\
47	0.00785516721127895\\
48	0.00706965049015106\\
49	0.00636268544113595\\
50	0.00572641689702236\\
51	0.00515377520732012\\
52	0.00463839768658811\\
53	0.0041745579179293\\
54	0.00375710212613637\\
55	0.00338139191352273\\
56	0.00304325272217046\\
57	0.00273892744995341\\
58	0.00246503470495807\\
59	0.00221853123446226\\
60	0.00199667811101604\\
61	0.00179701029991443\\
62	0.00161730926992299\\
63	0.00145557834293069\\
64	0.00131002050863762\\
65	0.00117901845777386\\
66	0.00106111661199647\\
67	0.000955004950796827\\
68	0.000859504455717144\\
69	0.00077355401014543\\
70	0.000696198609130887\\
71	0.000626578748217798\\
72	0.000563920873396018\\
73	0.000507528786056417\\
74	0.000456775907450775\\
75	0.000411098316705697\\
76	0.000369988485035128\\
77	0.000332989636531615\\
78	0.000299690672878453\\
79	0.000269721605590608\\
80	0.000242749445031547\\
81	0.000218474500528393\\
82	0.000196627050475553\\
83	0.000176964345427998\\
84	0.000159267910885198\\
85	0.000143341119796678\\
86	0.000129007007817011\\
87	0.000116106307035309\\
88	0.000104495676331779\\
89	9.40461086986007e-05\\
90	8.46414978287406e-05\\
91	7.61773480458666e-05\\
92	6.85596132412799e-05\\
93	6.17036519171519e-05\\
94	5.55332867254367e-05\\
95	4.99799580528931e-05\\
96	4.49819622476037e-05\\
97	4.04837660228434e-05\\
98	3.6435389420559e-05\\
99	3.27918504785031e-05\\
100	2.95126654306528e-05\\
};
\addlegendentry{$\varrho=0.9$}

\end{axis}
\end{tikzpicture}%
\definecolor{mycolor1}{rgb}{0.00000,0.44700,0.74100}%
\definecolor{mycolor2}{rgb}{0.85000,0.32500,0.09800}%
\begin{tikzpicture}

\begin{axis}[%
width=0.8\fwidth,
%height=0.33\fwidth,
at={(0\fwidth,0\fwidth)},
scale only axis,
xmin=0.9,
 tick label style={font=\scriptsize} , 
xlabel style={font=\color{white!15!black}},
xlabel={$ |i - j|$},
ymin=-0.03,
ymax=1.03,
xmax=60.9,
axis background/.style={fill=white},
axis x line*=bottom,
axis y line*=left,
xmajorgrids,
ymajorgrids,
legend style={legend cell align=left, align=left, draw=white!15!black,font = {\footnotesize}}
]
\addplot [color=mycolor1, line width=1.0pt]
  table[row sep=crcr]{%
1	1\\
2	0.6\\
3	0.279909897461042\\
4	0.179191691968152\\
5	0.130582584494419\\
6	0.102160790702494\\
7	0.0835958802077937\\
8	0.0705575360565466\\
9	0.0609189297267176\\
10	0.0535161041173487\\
11	0.0476596940834569\\
12	0.0429160059380057\\
13	0.0389988570952152\\
14	0.0357119498335893\\
15	0.0329162544711529\\
16	0.030510608231307\\
17	0.02841968562207\\
18	0.0265862715855896\\
19	0.0249661453600026\\
20	0.023524602258227\\
21	0.0222340334732084\\
22	0.0210722071117942\\
23	0.0200210247025744\\
24	0.0190656067415461\\
25	0.0181936101510325\\
26	0.0173947119282647\\
27	0.0166602136934231\\
28	0.0159827354072192\\
29	0.0153559756897033\\
30	0.014774522464039\\
31	0.014233702035832\\
32	0.0137294578199061\\
33	0.0132582521472478\\
34	0.0128169861942442\\
35	0.0124029342556563\\
36	0.0120136894559276\\
37	0.011647118646193\\
38	0.0113013247268024\\
39	0.0109746150098535\\
40	0.0106654745236044\\
41	0.0103725433827186\\
42	0.0100945975211468\\
43	0.00983053221990026\\
44	0.00957934796878718\\
45	0.00934013828593768\\
46	0.00911207918657557\\
47	0.00889442004676452\\
48	0.00868647565162925\\
49	0.00848761925303615\\
50	0.00829727649061813\\
51	0.00811492005367489\\
52	0.00794006498091103\\
53	0.00777226451100852\\
54	0.00761110641031416\\
55	0.00745620971496949\\
56	0.00730722183403267\\
57	0.00716381596786519\\
58	0.00702568880254918\\
59	0.00689255844657502\\
60	0.00676416258067001\\
61	0.00664025679556792\\
62	0.00652061309586271\\
63	0.00640501855094268\\
64	0.00629327407644289\\
65	0.00618519333174794\\
66	0.00608060172087881\\
67	0.00597933548565082\\
68	0.00588124088133203\\
69	0.00578617342619469\\
70	0.00569399721736175\\
71	0.00560458430622915\\
72	0.00551781412751059\\
73	0.00543357297662078\\
74	0.005351753530699\\
75	0.00527225440908868\\
76	0.00519497976954059\\
77	0.00511983893680419\\
78	0.00504674606062302\\
79	0.00497561980045912\\
80	0.00490638303454589\\
81	0.00483896259111162\\
82	0.00477328899983151\\
83	0.00470929626175782\\
84	0.0046469216361484\\
85	0.00458610544276626\\
86	0.00452679087835848\\
87	0.00446892384614477\\
88	0.00441245279725437\\
89	0.00435732858314795\\
90	0.0043035043181487\\
91	0.00425093525128544\\
92	0.00419957864672177\\
93	0.00414939367210883\\
94	0.00410034129425724\\
95	0.00405238418157564\\
96	0.00400548661277024\\
97	0.00395961439134286\\
98	0.00391473476546296\\
99	0.00387081635282473\\
100	0.00382782907013178\\
};
\addlegendentry{$\alpha=0.1$}

\addplot [color=mycolor2, line width=1.0pt]
  table[row sep=crcr]{%
1	1\\
2	0.6\\
3	0.243675718906871\\
4	0.143844618664973\\
5	0.0989630933079671\\
6	0.0740440635264012\\
7	0.0584190681067865\\
8	0.0478105564545639\\
9	0.040191504845111\\
10	0.0344854571981191\\
11	0.0300712340176363\\
12	0.0265668925759543\\
13	0.0237255140313178\\
14	0.0213808463423246\\
15	0.0194171195256723\\
16	0.017751400137267\\
17	0.0163228230618023\\
18	0.0150857759219386\\
19	0.0140054476243063\\
20	0.0130548404708158\\
21	0.0122127159461071\\
22	0.0114621521022782\\
23	0.0107895110792788\\
24	0.010183687490965\\
25	0.00963555281336067\\
26	0.00913753890583622\\
27	0.00868332183883882\\
28	0.00826757906691781\\
29	0.0078858009325314\\
30	0.00753414289846105\\
31	0.00720930865008676\\
32	0.00690845683493475\\
33	0.00662912607362388\\
34	0.0063691742195024\\
35	0.0061267288217439\\
36	0.00590014646559053\\
37	0.00568797919744227\\
38	0.00548894664284798\\
39	0.00530191272823423\\
40	0.00512586614807678\\
41	0.00495990389662176\\
42	0.00480321732057956\\
43	0.00465508025623756\\
44	0.00451483889839847\\
45	0.00438190311482817\\
46	0.0042557389725242\\
47	0.00413586228413967\\
48	0.00402183301663304\\
49	0.00391325043143463\\
50	0.00380974884749174\\
51	0.0037109939365319\\
52	0.00361667947459414\\
53	0.00352652448596463\\
54	0.0034402707256254\\
55	0.00335768045458432\\
56	0.00327853446932172\\
57	0.00320263035231843\\
58	0.00312978091542758\\
59	0.00305981281188265\\
60	0.00299256529613086\\
61	0.00292788911355236\\
62	0.00286564550456059\\
63	0.00280570530964968\\
64	0.0027479481637202\\
65	0.0026922617695243\\
66	0.00263854124136551\\
67	0.00258668851130059\\
68	0.002536611791049\\
69	0.00248822508364315\\
70	0.00244144773956853\\
71	0.00239620405276434\\
72	0.002352422892395\\
73	0.00231003736677345\\
74	0.00226898451622787\\
75	0.00222920503206239\\
76	0.00219064299907727\\
77	0.00215324565938994\\
78	0.00211696319554166\\
79	0.00208174853108833\\
80	0.00204755714706289\\
81	0.00201434691286383\\
82	0.00198207793027217\\
83	0.00195071238943026\\
84	0.00192021443573225\\
85	0.00189055004667978\\
86	0.00186168691784845\\
87	0.00183359435719325\\
88	0.00180624318699424\\
89	0.00177960565281002\\
90	0.00175365533886502\\
91	0.0017283670893497\\
92	0.00170371693516011\\
93	0.00167968202564591\\
94	0.00165624056497436\\
95	0.00163337175275241\\
96	0.00161105572858019\\
97	0.00158927352023743\\
98	0.00156800699522978\\
99	0.00154723881544529\\
100	0.00152695239469183\\
};
\addlegendentry{$\alpha=0.3$}

\end{axis}
\end{tikzpicture}%
}
\caption{\label{fig:Mij_models}
$(\M)_{ij}$ as a function of $|i - j |$ for Model 1 in \eqref{eq:AR1model} with various correlation parameters $\varrho$ (left), and for Model 2 in \eqref{eq:AR1model} with various decay parameters $\alpha$ and $\rho = 0.6$.
$p=100$. 
} 
\end{figure}

In {\bf Model~1}, $\M$ possesses an auto-regressive AR(1) structure:  
\beq \label{eq:AR1model}
(\M)_{ij} =  \varrho^{|i-j|}, 
\eeq 
where $|\varrho| \in [0,1)$.
When $\varrho \downarrow 0$, then $\M$ is close to an identity matrix scaled by $\eta$, and when $\varrho
\uparrow 1$, $\M$ tends to a singular matrix of rank 1.  
As illustrated in~\autoref{fig:Mij_models}, banding matrices allow for a good approximation, so all tapering-type estimators are computed with $\W (k)$ as \eqref{eq:Wband} in this subsection.  The optimal bandwidth $\hat k_o$ is chosen by consider the set of tapering matrices  
 $\mathbb{W} = \{ \bo W (k)  : k \in  [\![1,30]\!]\cup  [\![p-30,p]\!] \}$. 

\begin{figure}[!t]
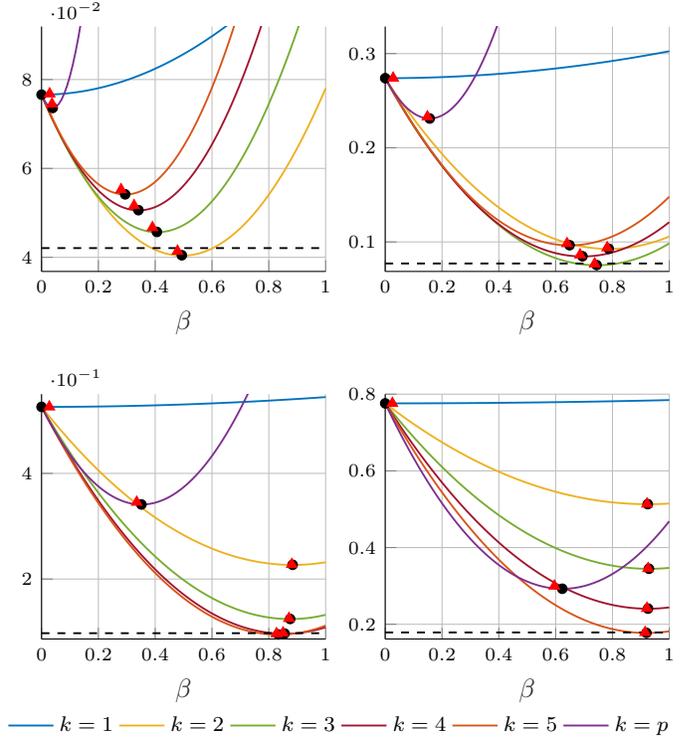

\centering
\setlength\fwidth{0.85\textwidth}
\subfloat{\input{tikz/vinf_n50_p100_rho0dot2_ell1.tex}}
\subfloat{\input{tikz/vinf_n50_p100_rho0dot4_ell1.tex}} \hspace{2pt}
\subfloat{\input{tikz/vinf_n50_p100_rho0dot6_ell1.tex}}
\subfloat{\input{tikz/vinf_n50_p100_rho0dot8_ell1.tex}}
\caption{NMSE  of \tabasco{} estimator using fixed $\W (k)$ as in \eqref{eq:Wband}. 
Samples are drawn from a MVN distribution, $\M$ as in Model 1 in \eqref{eq:AR1model} with $\varrho=0.2$ (top left),  $\varrho=0.4$ (top right), $\varrho=0.6$ (bottom left),  $\varrho=0.8$ (bottom right); $n=50$, $p=100$. 
The solid lines correspond to theoretical NMSE curves and the horizontal dashed line correspond to the NMSE obtained using $\hat k$. 
} \label{fig:AR1_beta}
\end{figure}

\autoref{fig:AR1_beta} illustrates a validation for the theoretical results:
it displays the theoretical normalized MSE (NMSE) curves, $L(\beta)=\E[ \| \hat \M_{\beta} - \M \|_{\Fr}^2]/\| \M \|_{\Fr}^2$  as a function of shrinkage parameter $\be$ for  \tabasco{} estimators using a fixed bandwidths $k \in [\![1,5 ]\!]$ and $k=p$ (i.e., $\W = \bo 1 \bo 1^\top$). 
In this setup, the data is generated from MVN distribution $\mathcal N_p(\bom\mu, \M)$ with $p=100$ and $n=50$ (similar results were obtained for others ES distributions and dimension setups). 
The black bullet ($\bullet$)  displays the theoretical minimum NMSE in \eqref{eq:MSEopt} attained for $\be_o\equiv \be_o(k)$ for each bandwidth $k$. 
The  empirical  average NMSE  for \tabasco{} using estimated $\hat \beta_0$ for each  fixed $k$ is displayed using red triangle ({\color{red} $\blacktriangle$}),  where the location on $\beta$ axis correspond to  empirical average $\hat \beta_0$.  
As can be noted from  \autoref{fig:AR1_beta}, \tabasco{} estimates the oracle shrinkage parameter $\beta_o$ very accurately since the black bullets and red triangles  are mostly overlapping for each bandwidth.
The dashed horizontal  line shows the average NMSE obtained by \tabasco{} when using the estimated optimal bandwidth $\hat k_0$.
One can notice that the optimal bandwidth selection using  \eqref{eq:hatk} is also accurate.
For example, in the case of   $\varrho=0.4$, the  optimal bandwidth is $k=3$ and  \tabasco{}  estimator attains an average NMSE that is very close to the theoretical minimum NMSE.    

\autoref{fig:AR_tdist} compares the performance of \tabasco{} with the state of the art in various setups.
The upper panel displays the NMSE curves  as a function of the sample size $n$ for four choices of correlation parameter $\varrho$ when the data follows a MVN distribution. 
The lower panel displays the same results when the data follows a MVT distribution with $\nu= 5$, which is heavy-tailed with marginal kurtosis $\mathrm{kurt}(x_i) = 6$ and elliptical kurtosis $\kappa =  \mathrm{kurt}(x_i)/3 = 2$. 
In the Gaussian case, all banding-type estimators outperform LWE thanks to the exploitation of the diagonally dominant structure of the covariance matrix.
In the heavy-tailed case, this is no longer true for STOA and ST-gaus, while ST-nong and \tabasco{} remain robust.
In all scenarios, \tabasco{} offers the lowest NMSE, and especially improves the performance when $n\ll p$. 

\begin{figure}[!t]
\setlength\fwidth{0.4\textwidth}
\begin{tikzpicture}

\begin{axis}[%
width=0.8\fwidth,
%height=0.33\fwidth,
%at={(0\fwidth,0\fwidth)},
scale only axis,
 tick label style={font=\footnotesize} , 
xmin=10,
xmax=115,
xlabel style={font=\color{white!15!black}},
xlabel={$n$},
ymin=-0.001,
ymax=1.001,
ylabel style={font=\color{white!15!black}},
ylabel={$\beta$},
axis background/.style={fill=white},
xmajorgrids,
ymajorgrids,
legend style={anchor=south east, legend cell align=left,  font = {\footnotesize}, anchor=south west, align=left, draw=none, legend columns=4,at={(-0.2,-0.37)}}
]
\addplot [color=red, dashdotted, line width=0.8pt, mark size=2.0pt, mark=triangle, mark options={solid, rotate=180, red}]
  table[row sep=crcr]{%
10	0.205388525215561\\
25	0.0975624147645011\\
40	0.0778526868764491\\
55	0.0767052263783963\\
70	0.0791140885601742\\
85	0.0856926145020936\\
100	0.0924756447151881\\
115	0.100056917592486\\
};
\addlegendentry{LWE, $\varrho=0.2 $}

\addplot [color=red, line width=0.8pt, mark size=2.0pt, mark=triangle, mark options={solid, rotate=180, red}]
  table[row sep=crcr]{%
10	0.225192396877919\\
25	0.15431415021047\\
40	0.168372825360852\\
55	0.196442945042742\\
70	0.225225580001624\\
85	0.254643678287651\\
100	0.282396239772343\\
115	0.30830278218007\\
};
\addlegendentry{$\varrho=0.4 $}

\addplot [color=red, dashed, line width=0.8pt, mark size=2.0pt, mark=triangle, mark options={solid, rotate=180, red}]
  table[row sep=crcr]{%
10	0.26933279153671\\
25	0.267678107195924\\
40	0.330175562803983\\
55	0.390504425803453\\
70	0.442822157359809\\
85	0.487493313268137\\
100	0.526112083716998\\
115	0.559304543400184\\
};
\addlegendentry{$\varrho=0.6 $}

\addplot [color=red, dotted, line width=0.8pt, mark size=2.0pt,mark=triangle, mark options={solid, rotate=180, red}]
  table[row sep=crcr]{%
10	0.376796823805654\\
25	0.482418334126833\\
40	0.582073618704297\\
55	0.652105062226239\\
70	0.702543830075915\\
85	0.740275724935368\\
100	0.769726785004595\\
115	0.793323217007636\\
};
\addlegendentry{$\varrho=0.8 $}

\addplot [color=blue, dashdotted, line width=0.8pt, mark size=2.0pt,mark=o, mark options={solid, blue}]
  table[row sep=crcr]{%
10	0.0128832796584193\\
25	0.0527794677998769\\
40	0.358787550751363\\
55	0.504743047735294\\
70	0.58436899283294\\
85	0.638773094835167\\
100	0.674810729761734\\
115	0.704061062995275\\
};
\addlegendentry{\tabasco{}, $\varrho=0.2 $}

\addplot [color=blue, line width=0.8pt, mark size=2.0pt, mark=o, mark options={solid, blue}]
  table[row sep=crcr]{%
10	0.0336089587109955\\
25	0.518705487086826\\
40	0.682966993282989\\
55	0.753961278893824\\
70	0.797889598037769\\
85	0.828443756638582\\
100	0.849601610009589\\
115	0.866643767653788\\
};
\addlegendentry{$\varrho=0.4 $}

\addplot [color=blue, dashed, line width=0.8pt, mark size=2.0pt,mark=o, mark options={solid, blue}]
  table[row sep=crcr]{%
10	0.15977960604657\\
25	0.71695904555693\\
40	0.807270287394314\\
55	0.849768181756655\\
70	0.874367804011666\\
85	0.89197470906287\\
100	0.904805662269229\\
115	0.915420642311718\\
};
\addlegendentry{$\varrho=0.6 $}

\addplot [color=blue, dotted, line width=0.8pt, mark size=2.0pt,mark=o, mark options={solid, blue}]
  table[row sep=crcr]{%
10	0.547125677995541\\
25	0.798502744292643\\
40	0.860089231773902\\
55	0.890395757897268\\
70	0.909564241249303\\
85	0.922381222925005\\
100	0.931778695678722\\
115	0.93914702789938\\
};
\addlegendentry{$\varrho=0.8 $}

\end{axis}

\end{tikzpicture}%
\caption{ Average  estimated shrinkage parameter $\beta$ for LWE and \tabasco{} when samples are from a MVN distribution, $\M$ has an AR(1) structure ($\varrho \in \{ 0.2, 0.4, 0.6, 0.8\}$) and $p = 100$. Banding matrices are used in \tabasco{}.} \label{fig:beta_gauss}
\end{figure}
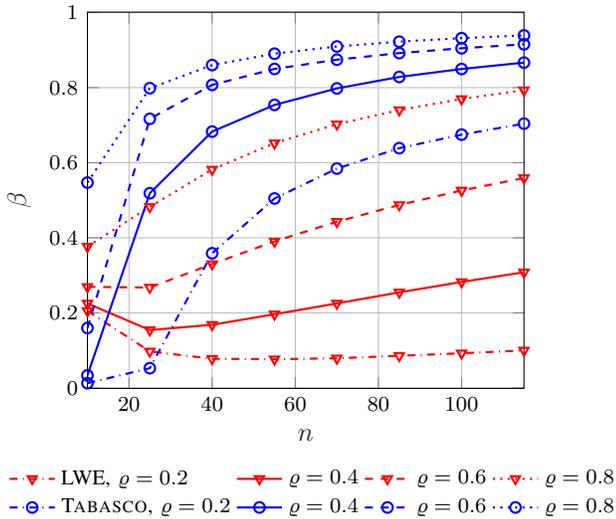

\autoref{fig:beta_gauss} displays the obtained (average) estimated shrinkage parameter $\hat \beta_o$ of \tabasco{} and LWE as a function of $n$. The average shrinkage parameter of \tabasco{}  is generally much larger than that of LWE. This means that it assign overall more weight on the banded SCM $\W \circ \S$ compared to LWE, which uses $\W(p)=\bo 1 \bo 1^\top$.  
This behavior is expected since banding the SCM should naturally improve the MSE when the true covariance matrix has a diagonally dominant structure. 

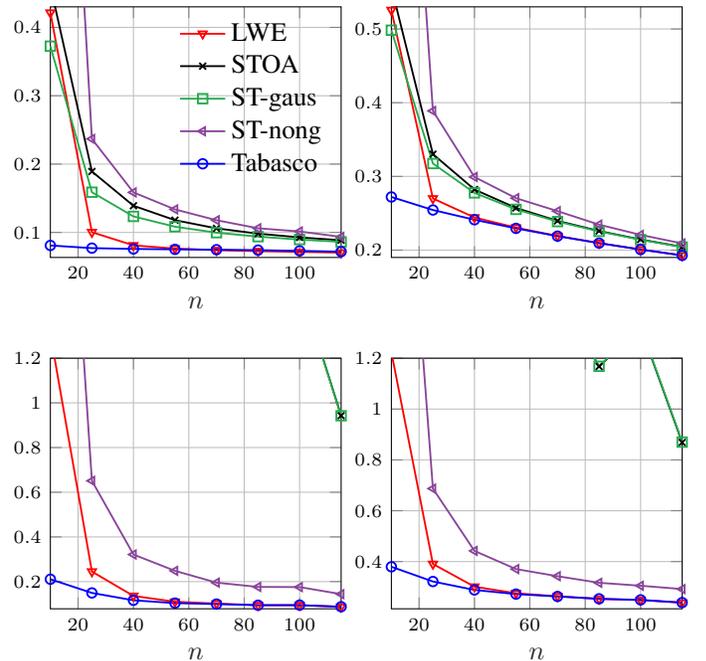
\begin{figure}[!t]
\centering
\setlength\fwidth{0.87\textwidth}
\subfloat{% This file was created by matlab2tikz.
\definecolor{mycolor1}{rgb}{0.14200,0.67200,0.30300}%
\definecolor{mycolor2}{rgb}{0.54200,0.27200,0.60300}%
\begin{tikzpicture}

\begin{axis}[%
width=0.245\fwidth,
%height=0.33\fwidth,
at={(0\fwidth,0\fwidth)},
scale only axis,
xmin=10,
xmax=115,
xtick={20,40,60,80,100},
ytick = {0.1,0.2,0.3,0.4,0.5}, 
xlabel style={font=\color{white!15!black}},
 tick label style={font=\scriptsize} , 
xlabel={$n$},
ymin=0.063384782546622,
ymax = 0.43,
ylabel style={font=\color{white!15!black}},
axis background/.style={fill=white},
xmajorgrids,
ymajorgrids,
legend style={legend cell align=left, align=left, fill=none, draw=none}
]
\addplot [color=red, line width=0.7pt, mark size=1.8pt, mark=triangle, mark options={solid, rotate=180, red}]
  table[row sep=crcr]{%
10	0.421501530205038\\
25	0.10036864499091\\
40	0.0810797685978442\\
55	0.0763854989774187\\
70	0.0740431035256935\\
85	0.0726066823377716\\
100	0.0715107818328131\\
115	0.0704809781832774\\
};
\addlegendentry{LWE}

\addplot [color=black,  line width=0.7pt, mark size=1.8pt,  mark=x, mark options={solid, black}]
  table[row sep=crcr]{%
10	0.460918999958143\\
25	0.189206083943737\\
40	0.139049421503074\\
55	0.117948918979298\\
70	0.105936119307902\\
85	0.0982092480767996\\
100	0.0927485759814686\\
115	0.0885535387774011\\
};
\addlegendentry{STOA}

\addplot [color=mycolor1,  line width=0.7pt, mark size=1.8pt,  mark=square, mark options={solid, mycolor1}]
  table[row sep=crcr]{%
10	0.372487298225597\\
25	0.158850497852085\\
40	0.123439910174653\\
55	0.108405580543926\\
70	0.0996386063328815\\
85	0.0938172261479216\\
100	0.0895593527483027\\
115	0.0862073028434835\\
};
\addlegendentry{ST-gaus}

\addplot [color=mycolor2,  line width=0.7pt, mark size=1.8pt,  mark=triangle, mark options={solid, rotate=90, mycolor2}]
  table[row sep=crcr]{%
10	1.2948172344377\\
25	0.236828074664148\\
40	0.158582898789788\\
55	0.133682972917569\\
70	0.117880292467105\\
85	0.106069374380344\\
100	0.101422322557861\\
115	0.093522337067972\\
};
\addlegendentry{ST-nong}

\addplot [color=blue,  line width=0.7pt, mark size=1.8pt,   mark=o, mark options={solid, blue}]
  table[row sep=crcr]{%
10	0.0809479405589102\\
25	0.0770275615306021\\
40	0.0758567540190079\\
55	0.0752431479313902\\
70	0.0747350309137317\\
85	0.0740947992423497\\
100	0.0730277495622928\\
115	0.0718881688664\\
};
\addlegendentry{Tabasco}

\end{axis}
\end{tikzpicture}%
}
\subfloat{% This file was created by matlab2tikz.
\definecolor{mycolor1}{rgb}{0.14200,0.67200,0.30300}%
\definecolor{mycolor2}{rgb}{0.54200,0.27200,0.60300}%
\begin{tikzpicture}

\begin{axis}[%
width=0.245\fwidth,
%height=0.33\fwidth,
at={(0\fwidth,0\fwidth)},
scale only axis,
xmin=10,
xmax=115,
xtick={20,40,60,80,100},
ytick = {0.1,0.2,0.3,0.4,0.5}, 
 tick label style={font=\scriptsize} , 
xlabel style={font=\color{white!15!black}},
xlabel={$n$},
ymin=0.19,
ymax = 0.53,
ylabel style={font=\color{white!15!black}},
axis background/.style={fill=white},
xmajorgrids,
ymajorgrids,
legend style={legend cell align=left, align=left, fill=none, draw=none}
]
\addplot [color=red, line width=0.7pt, mark size=1.8pt,  mark=triangle, mark options={solid, rotate=180, red}]
  table[row sep=crcr]{%
10	0.525678391861711\\
25	0.270010872024347\\
40	0.244213608237757\\
55	0.230477925392509\\
70	0.219198567726212\\
85	0.209521575709447\\
100	0.200812918154746\\
115	0.192799133373434\\
};
%\addlegendentry{LWE}

\addplot [color=black, line width=0.7pt, mark size=1.8pt, mark=x, mark options={solid, black}]
  table[row sep=crcr]{%
10	0.556691453608581\\
25	0.330153222163805\\
40	0.282208907404129\\
55	0.257190884298112\\
70	0.239659583150343\\
85	0.225947839628382\\
100	0.214383073880183\\
115	0.204402552918685\\
};
%\addlegendentry{STOA}

\addplot [color=mycolor1, line width=0.7pt, mark size=1.8pt, mark=square, mark options={solid, mycolor1}]
  table[row sep=crcr]{%
10	0.498322548502575\\
25	0.317451565358134\\
40	0.277625866872311\\
55	0.255243510942282\\
70	0.23866834896541\\
85	0.225397947911757\\
100	0.214068909364446\\
115	0.204155073160979\\
};
%\addlegendentry{ST-gaus}

\addplot [color=mycolor2, line width=0.7pt, mark size=1.8pt, mark=triangle, mark options={solid, rotate=90, mycolor2}]
  table[row sep=crcr]{%
10	1.21839152027721\\
25	0.38885013134313\\
40	0.299199512181395\\
55	0.270452565395207\\
70	0.252861125477686\\
85	0.234538841831489\\
100	0.220666896373482\\
115	0.208996682249144\\
};
%\addlegendentry{ST-nong}

\addplot [color=blue, line width=0.7pt, mark size=1.8pt, mark=o, mark options={solid, blue}]
  table[row sep=crcr]{%
10	0.272066681288488\\
25	0.254372574846419\\
40	0.241156641727964\\
55	0.229473105141576\\
70	0.21891410438731\\
85	0.209453102057669\\
100	0.200763587653361\\
115	0.19277439953861\\
};
%\addlegendentry{Tabasco}

\end{axis}
\end{tikzpicture}%
}  \hspace{2pt} 
\subfloat{% This file was created by matlab2tikz.
\definecolor{mycolor1}{rgb}{0.14200,0.67200,0.30300}%
\definecolor{mycolor2}{rgb}{0.54200,0.27200,0.60300}%
\begin{tikzpicture}

\begin{axis}[%
width=0.245\fwidth,
%height=0.33\fwidth,
at={(0\fwidth,0\fwidth)},
scale only axis,
xmin=10,
xmax=115,
xlabel style={font=\color{white!15!black}},
xlabel={$n$},
ymin=0.0783098628324136,
ymax=1.2,
ytick={0.2,0.4,0.6,0.8,1,1.2},
xtick={20,40,60,80,100},
 tick label style={font=\scriptsize} , 
ylabel style={font=\color{white!15!black}},
axis background/.style={fill=white},
xmajorgrids,
ymajorgrids,
legend style={legend cell align=left, align=left, fill=none, draw=none}
]
\addplot [color=red, line width=0.7pt, mark size=1.8pt,mark=triangle, mark options={solid, rotate=180, red}]
  table[row sep=crcr]{%
10	1.3084502351159\\
25	0.244902109705596\\
40	0.13589569073187\\
55	0.109838156684109\\
70	0.101631732999806\\
85	0.0945739863429331\\
100	0.094276713828129\\
115	0.0870109587026818\\
};
%\addlegendentry{LWE}

\addplot [color=black, line width=0.7pt, mark size=1.8pt, mark=x, mark options={solid, black}]
  table[row sep=crcr]{%
10	6.92223373316737\\
25	4.84648925625609\\
40	2.69053727061375\\
55	1.86727835198294\\
70	1.6866118114358\\
85	1.31185208007638\\
100	1.52979243276196\\
115	0.942751391359844\\
};
%\addlegendentry{STOA}

\addplot [color=mycolor1, line width=0.7pt, mark size=1.8pt,mark=square, mark options={solid, mycolor1}]
  table[row sep=crcr]{%
10	7.06391820105904\\
25	4.84908610179851\\
40	2.68832888288523\\
55	1.86504539708231\\
70	1.68537772338953\\
85	1.31081371319643\\
100	1.5296614477248\\
115	0.942476664097407\\
};
%\addlegendentry{ST-gaus}

\addplot [color=mycolor2, line width=0.7pt, mark size=1.8pt, mark=triangle, mark options={solid, rotate=90, mycolor2}]
  table[row sep=crcr]{%
10	3.43823135537358\\
25	0.651143366265818\\
40	0.321242679171014\\
55	0.248233628533583\\
70	0.195340692393526\\
85	0.176152773764788\\
100	0.175471604549573\\
115	0.143997145126325\\
};
%\addlegendentry{ST-nong}

\addplot [color=blue, line width=0.7pt, mark size=1.8pt, mark=o, mark options={solid, blue}]
  table[row sep=crcr]{%
10	0.210608852746009\\
25	0.149287138242854\\
40	0.116338270441239\\
55	0.103722283195788\\
70	0.0995074610438729\\
85	0.094253763903668\\
100	0.0946946314574597\\
115	0.0875569379507228\\
};
%\addlegendentry{Tabasco}

\end{axis}
\end{tikzpicture}%
}
\subfloat{% This file was created by matlab2tikz.
\definecolor{mycolor1}{rgb}{0.14200,0.67200,0.30300}%
\definecolor{mycolor2}{rgb}{0.54200,0.27200,0.60300}%
\begin{tikzpicture}

\begin{axis}[%
width=0.245\fwidth,
%height=0.33\fwidth,
at={(0\fwidth,0\fwidth)},
scale only axis,
xmin=10,
xmax=115,
xlabel style={font=\color{white!15!black}},
xlabel={$n$},
ymin=0.214333004144464,
ymax=1.2,
ytick={0.2,0.4,0.6,0.8,1,1.2},
xtick={20,40,60,80,100},
ylabel style={font=\color{white!15!black}},
 tick label style={font=\scriptsize} , 
axis background/.style={fill=white},
xmajorgrids,
ymajorgrids,
legend style={legend cell align=left, align=left, fill=none, draw=none}
]
\addplot [color=red, line width=0.7pt, mark size=1.8pt, mark=triangle, mark options={solid, rotate=180, red}]
  table[row sep=crcr]{%
10	1.22376773905966\\
25	0.390579774646408\\
40	0.30125425984278\\
55	0.275381589039403\\
70	0.263599488611472\\
85	0.253269569368323\\
100	0.248458283587163\\
115	0.238147782382738\\
};
%\addlegendentry{LWE}

\addplot [color=black, line width=0.7pt, mark size=1.8pt,mark=x, mark options={solid, black}]
  table[row sep=crcr]{%
10	5.63112189920143\\
25	3.92384176247116\\
40	2.266206224868\\
55	1.6137139725554\\
70	1.45235774644194\\
85	1.16750320514039\\
100	1.32021603351324\\
115	0.868437911023185\\
};
%\addlegendentry{STOA}

\addplot [color=mycolor1, line width=0.7pt, mark size=1.8pt, mark=square, mark options={solid, mycolor1}]
  table[row sep=crcr]{%
10	5.73819349017054\\
25	3.92794660024952\\
40	2.26723460950436\\
55	1.61139324318236\\
70	1.45356844282833\\
85	1.16815288553194\\
100	1.32217536695278\\
115	0.869999728380954\\
};
%\addlegendentry{ST-gaus}

\addplot [color=mycolor2, line width=0.7pt, mark size=1.8pt,mark=triangle, mark options={solid, rotate=90, mycolor2}]
  table[row sep=crcr]{%
10	2.85771741436131\\
25	0.687082019674252\\
40	0.442004307252907\\
55	0.37057240685064\\
70	0.342479017773145\\
85	0.316620172113531\\
100	0.304996781043299\\
115	0.291882044115955\\
};
%\addlegendentry{ST-nong}

\addplot [color=blue, line width=0.7pt, mark size=1.8pt, mark=o, mark options={solid, blue}]
  table[row sep=crcr]{%
10	0.379568235498162\\
25	0.321587986495897\\
40	0.288488584991755\\
55	0.272090123014699\\
70	0.262770309196803\\
85	0.253578272322906\\
100	0.249397915448769\\
115	0.238802395391235\\
};
%\addlegendentry{Tabasco}

\end{axis}
\end{tikzpicture}%
}
\vspace{-0.3cm} 
\caption{Average NMSE curves  when samples are from a MVN distribution (top row) and MVT distribution (bottom row) with $\nu=5$ d.o.f., $\M$ has a  permuted AR(1) structure with $\varrho = 0.2$ (left panel) and $\varrho= 0.4$ (right panel), and dimension is  $p = 100$.}
\label{fig:AR_gau_perm}
\end{figure}

\autoref{fig:AR_gau_perm} presents a comparison similar to \autoref{fig:AR_tdist} when the variables are permuted at random for each Monte Carlo trial, thus destroying the diagonally dominant structure of the AR(1) covariance matrix\footnote{Prominent algorithms for recovering hidden ordering-structure in the variables are the Best Permutation Analysis (BPA) \cite{rajaratnam2013best} or Isoband \cite{wagaman2009discovering}.
The perspective of their joint use with \tabasco{} is left for further studies.}.
The hypothesis is that any banding estimator with optimal bandwidth selection should be able to select the bandwidth $k=p$ accordingly.  
Note that LWE is invariant to variable permutations, and hence its  results stays the same for both of these scenarios. 
In this setup, \tabasco{} performs better  that LWE for $n \ll p$ and  equally well as LWE  for  $n$ large enough.  This result implies that bandwidth selection of \tabasco{} is consistent: it chooses  $k=p$ since the true covariance matrix does not have a diagonally dominant structure. 
The improvement brought at low sample support can be explained by the fact that an ES distribution is assumed by \tabasco{}, which allows for a better estimation of the oracle parameter (LWE only assumes finite 4th order moments).
This  example confirms that \tabasco{} always benefits from banding and bandwidth selection: it offers significantly improved NMSE compared to RSCM when banding structure is present in the covariance matrix, while it does not perform worse when such structure does not exist, thanks to its robust and efficient bandwidth selection.

\begin{figure*}[!t]
\centering
\setlength\fwidth{0.87\textwidth}
\subfloat{% This file was created by matlab2tikz.
\definecolor{mycolor1}{rgb}{0.14200,0.67200,0.30300}%
\definecolor{mycolor2}{rgb}{0.54200,0.27200,0.60300}%
\begin{tikzpicture}

\begin{axis}[%
width=0.245\fwidth,
%height=0.33\fwidth,
at={(0\fwidth,0\fwidth)},
scale only axis,
xmin=10,
xmax=115,
xtick={20,40,60,80,100},
xlabel style={font=\color{white!15!black}},
xlabel={$n$},
 tick label style={font=\scriptsize} , 
ymin=0.0208964487986457,
ymax = 0.5,
ytick = {0.1,0.2,0.3,0.4,0.5}, 
ylabel style={font=\color{white!15!black}},
axis background/.style={fill=white},
xmajorgrids,
ymajorgrids,
title={$\varrho=0.2 $},
legend style={legend cell align=left, align=left, fill=none, draw=none}
]
\addplot [color=red,  line width=0.7pt, mark size=1.8pt,    mark=triangle, mark options={solid, rotate=180, red}]
  table[row sep=crcr]{%
10	0.421388205395662\\
25	0.100309573880561\\
40	0.0810018864340239\\
55	0.0763101819514599\\
70	0.0740613126315062\\
85	0.0726242596510709\\
100	0.0714801790459146\\
115	0.0704677042441133\\
};
\addlegendentry{LWE}

\addplot [color=black,  line width=0.7pt, mark size=1.8pt,   mark=x, mark options={solid, black}]
  table[row sep=crcr]{%
10	0.46122396269448\\
25	0.189203597211247\\
40	0.139131309105173\\
55	0.117533248591795\\
70	0.100021918992804\\
85	0.081785686320972\\
100	0.0651498347426259\\
115	0.0543141153824146\\
};
\addlegendentry{STOA}

\addplot [color=mycolor1,  line width=0.7pt, mark size=1.8pt, mark=square, mark options={solid, mycolor1}]
  table[row sep=crcr]{%
10	0.372925223053353\\
25	0.164187678261226\\
40	0.102298761728827\\
55	0.0736766354018533\\
70	0.0581237312562378\\
85	0.0482655290112879\\
100	0.0412008472481337\\
115	0.0362929749221009\\
};
\addlegendentry{ST-gaus}

\addplot [color=mycolor2,  line width=0.7pt, mark size=1.8pt,    mark=triangle, mark options={solid, rotate=90, mycolor2}]
  table[row sep=crcr]{%
10	1.2694911990326\\
25	0.249213527716151\\
40	0.145655795409103\\
55	0.0935374829064041\\
70	0.0720098464963754\\
85	0.0613352604491353\\
100	0.0514994561385525\\
115	0.0466581652863365\\
};
\addlegendentry{ST-nong}

\addplot [color=blue,  line width=0.7pt, mark size=1.8pt,    mark=o, mark options={solid, blue}]
  table[row sep=crcr]{%
10	0.0809793367466806\\
25	0.0740455197793713\\
40	0.04860460699498\\
55	0.0400309298363954\\
70	0.0350667366553499\\
85	0.0313702480198924\\
100	0.0283812666224246\\
115	0.0261205609983072\\
};
\addlegendentry{\tabasco{}}

\end{axis}
\end{tikzpicture}%
}
\subfloat{% This file was created by matlab2tikz.
\definecolor{mycolor1}{rgb}{0.14200,0.67200,0.30300}%
\definecolor{mycolor2}{rgb}{0.54200,0.27200,0.60300}%
\begin{tikzpicture}

\begin{axis}[%
width=0.245\fwidth,
%height=0.33\fwidth,
at={(0\fwidth,0\fwidth)},
scale only axis,
xmin=10,
xmax=115,
xtick={20,40,60,80,100},
ytick = {0.1,0.2,0.3,0.4,0.5}, 
 tick label style={font=\scriptsize} , 
xlabel style={font=\color{white!15!black}},
xlabel={$n$},
ymin=0.0336644869184433,
ymax = 0.54,
ylabel style={font=\color{white!15!black}},
axis background/.style={fill=white},
xmajorgrids,
ymajorgrids,
title={$\varrho=0.4 $}, 
legend style={legend cell align=left, align=left, fill=none, draw=none}
]
\addplot [color=red, line width=0.7pt, mark size=1.8pt, mark=triangle, mark options={solid, rotate=180, red}]
  table[row sep=crcr]{%
10	0.525513131146538\\
25	0.270047706765539\\
40	0.24416621658863\\
55	0.230518804566445\\
70	0.219322545728277\\
85	0.209466895014301\\
100	0.200877587837181\\
115	0.192745493873973\\
};
%\addlegendentry{LWE}

\addplot [color=black, line width=0.7pt, mark size=1.8pt,  mark=x, mark options={solid, black}]
  table[row sep=crcr]{%
10	0.557575387568379\\
25	0.285436351582111\\
40	0.164197149121585\\
55	0.124187792793014\\
70	0.103996872678986\\
85	0.0909036298985663\\
100	0.0804788600882667\\
115	0.0719250228247473\\
};
%\addlegendentry{STOA}

\addplot [color=mycolor1, line width=0.7pt, mark size=1.8pt,  mark=square, mark options={solid, mycolor1}]
  table[row sep=crcr]{%
10	0.474321062301439\\
25	0.180414542396545\\
40	0.12672553953853\\
55	0.0947555060942857\\
70	0.0741051576761534\\
85	0.0618891525273407\\
100	0.0532060807131955\\
115	0.0472645494532335\\
};
%\addlegendentry{ST-gaus}

\addplot [color=mycolor2, line width=0.7pt, mark size=1.8pt,  mark=triangle, mark options={solid, rotate=90, mycolor2}]
  table[row sep=crcr]{%
10	1.16526678522785\\
25	0.263800194428618\\
40	0.168510183465771\\
55	0.115044814060003\\
70	0.0865015355206526\\
85	0.0720313335985859\\
100	0.0614319225726784\\
115	0.0563209716321833\\
};
%\addlegendentry{ST-nong}

\addplot [color=blue, line width=0.7pt, mark size=1.8pt, mark=o, mark options={solid, blue}]
  table[row sep=crcr]{%
10	0.272040941511234\\
25	0.125086312346454\\
40	0.0902870365334872\\
55	0.0723835363718454\\
70	0.0606657872925614\\
85	0.052479569072472\\
100	0.0464513618769075\\
115	0.0420806086480541\\
};
%\addlegendentry{Tabasco}

\end{axis}
\end{tikzpicture}%
}
\subfloat{% This file was created by matlab2tikz.
\definecolor{mycolor1}{rgb}{0.14200,0.67200,0.30300}%
\definecolor{mycolor2}{rgb}{0.54200,0.27200,0.60300}%
\begin{tikzpicture}

\begin{axis}[%
width=0.245\fwidth,
%height=0.33\fwidth,
at={(0\fwidth,0\fwidth)},
scale only axis,
xmin=10,
xmax=115,
 tick label style={font=\scriptsize} , 
xlabel style={font=\color{white!15!black}},
xlabel={$n$},
ymin=0.0507699442486262,
ymax = 0.5,
ylabel style={font=\color{white!15!black}},
axis background/.style={fill=white},
xmajorgrids,
ymajorgrids,
xtick={20,40,60,80,100},
ytick = {0.1,0.2,0.3,0.4,0.5}, 
title={$\varrho=0.6 $}, 
legend style={legend cell align=left, align=left, fill=none, draw=none}
]
\addplot [color=red, line width=0.7pt, mark size=1.8pt, mark=triangle, mark options={solid, rotate=180, red}]
  table[row sep=crcr]{%
10	0.634751718672683\\
25	0.428030850697016\\
40	0.372917087754849\\
55	0.333470949181421\\
70	0.301844391185302\\
85	0.275950588896406\\
100	0.254039128609047\\
115	0.235642012327365\\
};
%\addlegendentry{LWE}

\addplot [color=black, line width=0.7pt, mark size=1.8pt, mark=x, mark options={solid, black}]
  table[row sep=crcr]{%
10	0.644431143446459\\
25	0.286302390901258\\
40	0.184981813864624\\
55	0.141060809220168\\
70	0.114950918587138\\
85	0.0968569708771548\\
100	0.0830734567845264\\
115	0.073408128064207\\
};
%\addlegendentry{STOA}

\addplot [color=mycolor1, line width=0.7pt, mark size=1.8pt, mark=square, mark options={solid, mycolor1}]
  table[row sep=crcr]{%
10	0.488559546609828\\
25	0.20835214725371\\
40	0.135831995573022\\
55	0.10489010962313\\
70	0.0852986995186614\\
85	0.0717714545530481\\
100	0.0615646089824303\\
115	0.0548520446560123\\
};
%\addlegendentry{ST-gaus}

\addplot [color=mycolor2, line width=0.7pt, mark size=1.8pt, mark=triangle, mark options={solid, rotate=90, mycolor2}]
  table[row sep=crcr]{%
10	0.941932239832303\\
25	0.251349430601018\\
40	0.154948791064674\\
55	0.116270658691382\\
70	0.0943896051251522\\
85	0.0787901253565034\\
100	0.0667286821231975\\
115	0.0594190353821245\\
};
%\addlegendentry{ST-nong}

\addplot [color=blue, line width=0.7pt, mark size=1.8pt,mark=o, mark options={solid, blue}]
  table[row sep=crcr]{%
10	0.435001912679731\\
25	0.158936665917267\\
40	0.114616998480798\\
55	0.0913956352976611\\
70	0.0757400421345105\\
85	0.0648408231178333\\
100	0.0563687559448844\\
115	0.0507699442486262\\
};
%\addlegendentry{Tabasco}

\end{axis}
\end{tikzpicture}%
} 
\subfloat{% This file was created by matlab2tikz.
\definecolor{mycolor1}{rgb}{0.14200,0.67200,0.30300}%
\definecolor{mycolor2}{rgb}{0.54200,0.27200,0.60300}%
\begin{tikzpicture}

\begin{axis}[%
width=0.245\fwidth,
%height=0.33\fwidth,
at={(0\fwidth,0\fwidth)},
scale only axis,
xmin=10,
xmax=115,
 tick label style={font=\scriptsize} , 
xlabel style={font=\color{white!15!black}},
xlabel={$n$},
ymin=0.0552820090141459,
ymax = 0.5,
ylabel style={font=\color{white!15!black}},
xtick={20,40,60,80,100},
ytick = {0.1,0.2,0.3,0.4,0.5}, 
axis background/.style={fill=white},
xmajorgrids,
ymajorgrids,
title={$\varrho=0.8 $}, 
legend style={legend cell align=left, align=left, fill=none, draw=none}
]
\addplot [color=red, line width=0.7pt, mark size=1.8pt,  mark=triangle, mark options={solid, rotate=180, red}]
  table[row sep=crcr]{%
10	0.659799980875326\\
25	0.439371236804732\\
40	0.341694026820617\\
55	0.280642088964092\\
70	0.237359881246668\\
85	0.206181538232452\\
100	0.181913093980105\\
115	0.163388815902562\\
};
%\addlegendentry{LWE}

\addplot [color=black, line width=0.7pt, mark size=1.8pt,  mark=x, mark options={solid, black}]
  table[row sep=crcr]{%
10	0.623226991347306\\
25	0.278229301154114\\
40	0.181655996164373\\
55	0.13715278098867\\
70	0.10980158141388\\
85	0.0924380982738823\\
100	0.0798659432594798\\
115	0.0709101538235731\\
};
%\addlegendentry{STOA}

\addplot [color=mycolor1, line width=0.7pt, mark size=1.8pt, mark=square, mark options={solid, mycolor1}]
  table[row sep=crcr]{%
10	0.487925691517437\\
25	0.216623527194569\\
40	0.144688415132019\\
55	0.110186754326991\\
70	0.0890717650009035\\
85	0.0753972382997714\\
100	0.0651469442171782\\
115	0.0582182806059913\\
};
%\addlegendentry{ST-gaus}

\addplot [color=mycolor2, line width=0.7pt, mark size=1.8pt,  mark=triangle, mark options={solid, rotate=90, mycolor2}]
  table[row sep=crcr]{%
10	0.684092532475137\\
25	0.234972311202957\\
40	0.151855605011749\\
55	0.115848797426424\\
70	0.0930034618108073\\
85	0.0783873146981564\\
100	0.0673014483477707\\
115	0.0602201535303051\\
};
%\addlegendentry{ST-nong}

\addplot [color=blue, line width=0.7pt, mark size=1.8pt,  mark=o, mark options={solid, blue}]
  table[row sep=crcr]{%
10	0.355526229877736\\
25	0.180543062335463\\
40	0.127495342442498\\
55	0.100034948381937\\
70	0.0823091045347243\\
85	0.0705659673765752\\
100	0.0614528551689625\\
115	0.0552820090141459\\
};
%\addlegendentry{Tabasco}

\end{axis}
\end{tikzpicture}%
} \hspace{2pt} 
\subfloat{% This file was created by matlab2tikz.
\definecolor{mycolor1}{rgb}{0.14200,0.67200,0.30300}%
\definecolor{mycolor2}{rgb}{0.54200,0.27200,0.60300}%
\begin{tikzpicture}

\begin{axis}[%
width=0.245\fwidth,
%height=0.33\fwidth,
at={(0\fwidth,0\fwidth)},
scale only axis,
xmin=10,
xmax=115,
xtick={50,100,150},
ytick={0.2,0.4,0.6,0.8,1,1.2},
xtick={20,40,60,80,100},
 tick label style={font=\scriptsize} , 
xlabel style={font=\color{white!15!black}},
xlabel={$n$},
ymin=0.0502391697770027,
ymax=1.2,
ylabel style={font=\color{white!15!black}},
axis background/.style={fill=white},
xmajorgrids,
ymajorgrids,
legend style={legend cell align=left, align=left, fill=none, draw=none}
]
\addplot [color=red, line width=0.7pt, mark size=1.8pt,  mark=triangle, mark options={solid, rotate=180, red}]
  table[row sep=crcr]{%
10	1.78238535903651\\
25	0.247660744261288\\
40	0.141400145436929\\
55	0.108225032513963\\
70	0.101014895925162\\
85	0.0949352705669861\\
100	0.0923081291905833\\
115	0.089480521633637\\
};
%\addlegendentry{LWE}

\addplot [color=black, line width=0.7pt, mark size=1.8pt, mark=x, mark options={solid, black}]
  table[row sep=crcr]{%
10	11.3503899950374\\
25	4.89455877526661\\
40	3.07634499073816\\
55	1.74204202008161\\
70	1.50725051211254\\
85	1.29127995782966\\
100	1.27828616105313\\
115	1.12822444431713\\
};
%\addlegendentry{STOA}

\addplot [color=mycolor1, line width=0.7pt, mark size=1.8pt,  mark=square, mark options={solid, mycolor1}]
  table[row sep=crcr]{%
10	11.6809691990806\\
25	4.89960024937417\\
40	3.07141428358584\\
55	1.73074653023981\\
70	1.49224300291374\\
85	1.27666502384632\\
100	1.26609376221559\\
115	1.11861542110709\\
};
%\addlegendentry{ST-gaus}

\addplot [color=mycolor2, line width=0.7pt, mark size=1.8pt,  mark=triangle, mark options={solid, rotate=90, mycolor2}]
  table[row sep=crcr]{%
10	4.56564487650056\\
25	0.629744632181775\\
40	0.334349825130945\\
55	0.222564550067087\\
70	0.17952959342376\\
85	0.149378058932662\\
100	0.134096115739603\\
115	0.118083963413918\\
};
%\addlegendentry{ST-nong}

\addplot [color=blue,line width=0.7pt, mark size=1.8pt,  mark=o, mark options={solid, blue}]
  table[row sep=crcr]{%
10	0.27272553896473\\
25	0.151842081235255\\
40	0.102465641505196\\
55	0.0787035202197611\\
70	0.0719976446526393\\
85	0.0645741917345378\\
100	0.0606877127848368\\
115	0.0558212997522252\\
};
%\addlegendentry{Tabasco}

\end{axis}
\end{tikzpicture}%
}
\subfloat{% This file was created by matlab2tikz.
\definecolor{mycolor1}{rgb}{0.14200,0.67200,0.30300}%
\definecolor{mycolor2}{rgb}{0.54200,0.27200,0.60300}%
\begin{tikzpicture}

\begin{axis}[%
width=0.245\fwidth,
%height=0.33\fwidth,
at={(0\fwidth,0\fwidth)},
scale only axis,
xmin=10,
xmax=115,
ytick={0.2,0.4,0.6,0.8,1,1.2},
xtick={20,40,60,80,100},
xlabel style={font=\color{white!15!black}},
xlabel={$n$},
ymin=0.0648603476800369,
ymax=1.2,
 tick label style={font=\scriptsize} , 
ylabel style={font=\color{white!15!black}},
axis background/.style={fill=white},
xmajorgrids,
ymajorgrids,
legend style={legend cell align=left, align=left, fill=none, draw=none}
]
\addplot [color=red, line width=0.7pt, mark size=1.8pt, mark=triangle, mark options={solid, rotate=180, red}]
  table[row sep=crcr]{%
10	1.58846608157297\\
25	0.394946558086063\\
40	0.305963868015247\\
55	0.274249900609795\\
70	0.263543341955782\\
85	0.253476508096269\\
100	0.246998506653678\\
115	0.240367265006693\\
};
%\addlegendentry{LWE}

\addplot [color=black, line width=0.7pt, mark size=1.8pt, mark=x, mark options={solid, black}]
  table[row sep=crcr]{%
10	9.09887376833166\\
25	4.03819657818215\\
40	2.50282637019721\\
55	1.43667781331262\\
70	1.22915950293766\\
85	1.0436781843147\\
100	1.02761853318782\\
115	0.931677759860324\\
};
%\addlegendentry{STOA}

\addplot [color=mycolor1, line width=0.7pt, mark size=1.8pt, mark=square, mark options={solid, mycolor1}]
  table[row sep=crcr]{%
10	9.27627385651063\\
25	3.98769474097123\\
40	2.46987375284543\\
55	1.42252194670749\\
70	1.22448501847584\\
85	1.03998496249385\\
100	1.0235798103343\\
115	0.926918447822008\\
};
%\addlegendentry{ST-gaus}

\addplot [color=mycolor2, line width=0.7pt, mark size=1.8pt, mark=triangle, mark options={solid, rotate=90, mycolor2}]
  table[row sep=crcr]{%
10	3.74557942097376\\
25	0.631119743117498\\
40	0.339949993130694\\
55	0.243391962936299\\
70	0.202678673983778\\
85	0.170354553999268\\
100	0.150805156009978\\
115	0.133048290960338\\
};
%\addlegendentry{ST-nong}

\addplot [color=blue, line width=0.7pt, mark size=1.8pt, mark=o, mark options={solid, blue}]
  table[row sep=crcr]{%
10	0.427367743129919\\
25	0.229617575542525\\
40	0.175212434976903\\
55	0.138954674342316\\
70	0.122746394917085\\
85	0.108091074105377\\
100	0.09978164867633\\
115	0.0906639698489475\\
};
%\addlegendentry{Tabasco}

\end{axis}
\end{tikzpicture}%
}
\subfloat{% This file was created by matlab2tikz.
\definecolor{mycolor1}{rgb}{0.14200,0.67200,0.30300}%
\definecolor{mycolor2}{rgb}{0.54200,0.27200,0.60300}%
\begin{tikzpicture}

\begin{axis}[%
width=0.245\fwidth,
%height=0.33\fwidth,
at={(0\fwidth,0\fwidth)},
scale only axis,
xmin=10,
xmax=115,
xtick={20,40,60,80,100},
xlabel style={font=\color{white!15!black}},
xlabel={$n$},
 tick label style={font=\scriptsize} , 
ymin=0.0996210632090237,
ytick={0.2,0.4,0.6,0.8,1,1.2},
ymax=1.2,
ylabel style={font=\color{white!15!black}},
axis background/.style={fill=white},
xmajorgrids,
ymajorgrids,
legend style={legend cell align=left, align=left, fill=none, draw=none}
]
\addplot [color=red, line width=0.7pt, mark size=1.8pt,  mark=triangle, mark options={solid, rotate=180, red}]
  table[row sep=crcr]{%
10	1.34222570333092\\
25	0.54575303483332\\
40	0.463618059994889\\
55	0.421246522294391\\
70	0.396840360020582\\
85	0.374174822532173\\
100	0.357046574054364\\
115	0.340212776411591\\
};
%\addlegendentry{LWE}

\addplot [color=black,  line width=0.7pt, mark size=1.8pt,  mark=x, mark options={solid, black}]
  table[row sep=crcr]{%
10	6.34524670801555\\
25	2.84811761943188\\
40	1.73064379042305\\
55	1.03157715398456\\
70	0.878232641637688\\
85	0.735584037357028\\
100	0.714520406958265\\
115	0.65960018272793\\
};
%\addlegendentry{STOA}

\addplot [color=mycolor1, line width=0.7pt, mark size=1.8pt,   mark=square, mark options={solid, mycolor1}]
  table[row sep=crcr]{%
10	6.28044032357112\\
25	2.78342979968431\\
40	1.6981374395043\\
55	1.01352099735484\\
70	0.866749151340924\\
85	0.729297160509844\\
100	0.710540543824973\\
115	0.657422235115295\\
};
%\addlegendentry{ST-gaus}

\addplot [color=mycolor2,  line width=0.7pt, mark size=1.8pt,  mark=triangle, mark options={solid, rotate=90, mycolor2}]
  table[row sep=crcr]{%
10	2.65627879287611\\
25	0.567576696499149\\
40	0.342217929192866\\
55	0.244273440966429\\
70	0.204028307261086\\
85	0.173342022644699\\
100	0.158704585218494\\
115	0.140695092578489\\
};
%\addlegendentry{ST-nong}

\addplot [color=blue,  line width=0.7pt, mark size=1.8pt,  mark=o, mark options={solid, blue}]
  table[row sep=crcr]{%
10	0.573345341327857\\
25	0.292256655965451\\
40	0.21984909039556\\
55	0.173610066553108\\
70	0.151786965203496\\
85	0.132812297774009\\
100	0.122208350436066\\
115	0.110690070232249\\
};
%\addlegendentry{Tabasco}

\end{axis}
\end{tikzpicture}%
}
\subfloat{% This file was created by matlab2tikz.
\definecolor{mycolor1}{rgb}{0.14200,0.67200,0.30300}%
\definecolor{mycolor2}{rgb}{0.54200,0.27200,0.60300}%
\begin{tikzpicture}

\begin{axis}[%
width=0.245\fwidth,
%height=0.33\fwidth,
at={(0\fwidth,0\fwidth)},
scale only axis,
xmin=10,
xmax=115,
 tick label style={font=\scriptsize} , 
ytick={0.2,0.4,0.6,0.8,1,1.2},
xtick={20,40,60,80,100},
xlabel style={font=\color{white!15!black}},
xlabel={$n$},
ymin=0.10656820293731,
ymax=1.2,
ylabel style={font=\color{white!15!black}},
axis background/.style={fill=white},
xmajorgrids,
ymajorgrids,
legend style={legend cell align=left, align=left, fill=none, draw=none}
]
\addplot [color=red, line width=0.7pt, mark size=1.8pt,mark=triangle, mark options={solid, rotate=180, red}]
  table[row sep=crcr]{%
10	1.04901769274967\\
25	0.580285713258883\\
40	0.483369294856306\\
55	0.417800954692053\\
70	0.375937238814948\\
85	0.338762757501303\\
100	0.313303435917393\\
115	0.289223299485478\\
};
%\addlegendentry{LWE}

\addplot [color=black, line width=0.7pt, mark size=1.8pt, mark=x, mark options={solid, black}]
  table[row sep=crcr]{%
10	3.64447121979474\\
25	1.54581184820271\\
40	0.974485372543345\\
55	0.613299092092512\\
70	0.511424418076176\\
85	0.419612003781099\\
100	0.401625538956648\\
115	0.377311789968813\\
};
%\addlegendentry{STOA}

\addplot [color=mycolor1, line width=0.7pt, mark size=1.8pt,mark=square, mark options={solid, mycolor1}]
  table[row sep=crcr]{%
10	3.37526806008386\\
25	1.46277940349431\\
40	0.938906541941594\\
55	0.596455285426831\\
70	0.501780183549155\\
85	0.414594512834484\\
100	0.398638841511323\\
115	0.375641877043801\\
};
%\addlegendentry{ST-gaus}

\addplot [color=mycolor2, line width=0.7pt, mark size=1.8pt,mark=triangle, mark options={solid, rotate=90, mycolor2}]
  table[row sep=crcr]{%
10	1.61475694497524\\
25	0.465609334456726\\
40	0.318713708784133\\
55	0.238049820781156\\
70	0.203931720006165\\
85	0.171546171949723\\
100	0.154495569024332\\
115	0.139399958024391\\
};
%\addlegendentry{ST-nong}

\addplot [color=blue, line width=0.7pt, mark size=1.8pt,mark=o, mark options={solid, blue}]
  table[row sep=crcr]{%
10	0.553093237270405\\
25	0.322205438924852\\
40	0.244042835294069\\
55	0.190191432452749\\
70	0.164510430443775\\
85	0.141900929583793\\
100	0.130506218478823\\
115	0.118409114374789\\
};
%\addlegendentry{Tabasco}

\end{axis}
\end{tikzpicture}%
}
\vspace{-0.2cm} 
\caption{Average NMSE curves  when samples are from a MVN distribution (upper panel)  and  $t$-distribution with $\nu=5$ d.o.f.  (lower panel), $\M$ has an AR(1) structure with $\varrho \in \{ 0.2, 0.4, 0.6,0.8\}$ from left to right.  Dimension is  $p = 100$ and banding matrices are used in STOA, ST-gaus, ST-nong and \tabasco{}.} \label{fig:AR_tdist}
\end{figure*}
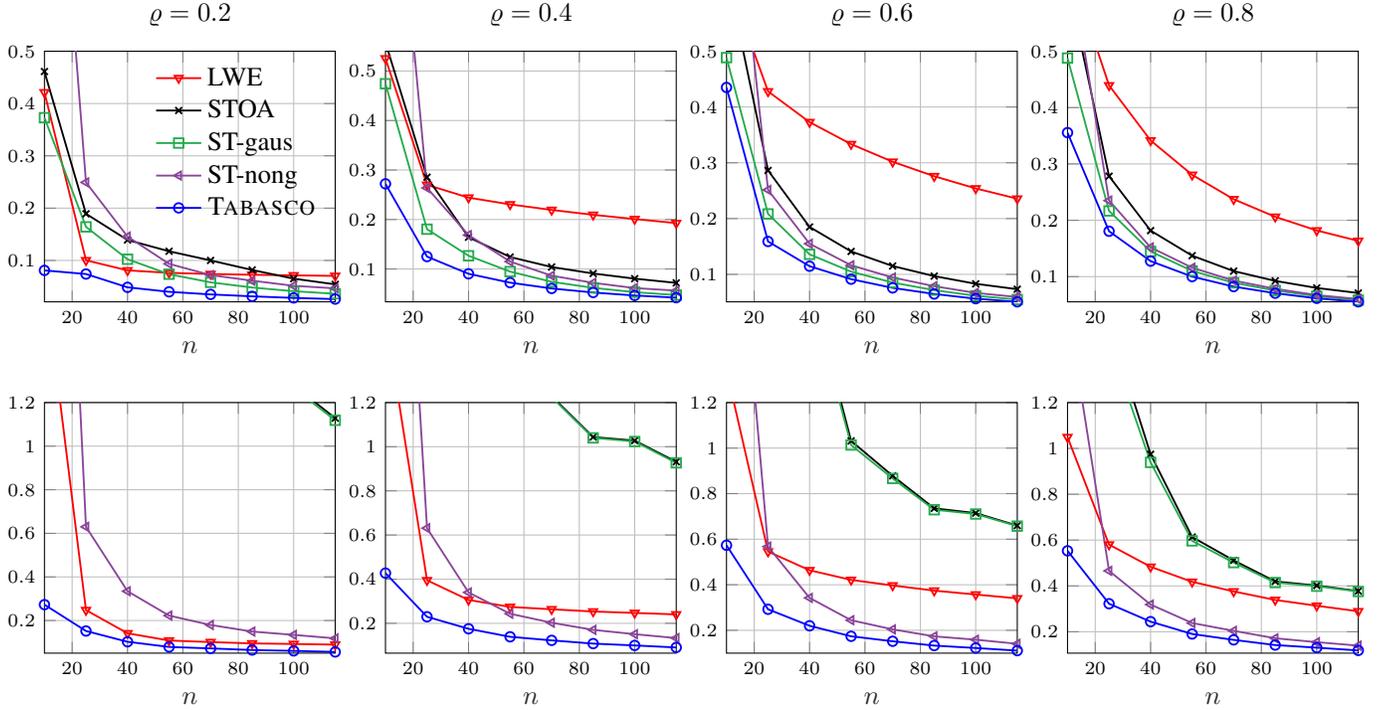

\subsection{Model 2} 

\begin{figure}[!t]
\centering
\setlength\fwidth{0.82\textwidth}
\subfloat{% This file was created by matlab2tikz.
\definecolor{mycolor1}{rgb}{0.14200,0.67200,0.30300}%
\definecolor{mycolor2}{rgb}{0.54200,0.27200,0.60300}%
\begin{tikzpicture}

\begin{axis}[%
width=0.245\fwidth,
%height=0.33\fwidth,
at={(0\fwidth,0\fwidth)},
scale only axis,
xmin=50,
xmax=350,
 tick label style={font=\scriptsize} , 
xlabel style={font=\color{white!15!black}},
yticklabel=\pgfkeys{/pgf/number format/.cd,fixed,precision=2,zerofill}\pgfmathprintnumber{\tick},
 ytick={0.06,0.08,0.1,0.12,0.14,0.16},
xlabel={$n$},
ymin=0.0405663718172311,
ymax=0.164255042692645,
ylabel style={font=\color{white!15!black}},
axis background/.style={fill=white},
xmajorgrids,
ymajorgrids,
legend style={legend cell align=left, align=left, fill=none, draw=none}
]
\addplot [color=blue, line width=0.7pt,  mark size=1.8pt,  mark=o, mark options={solid, blue}]
  table[row sep=crcr]{%
50	0.125563322039169\\
100	0.0846914830676173\\
150	0.0666701387984449\\
200	0.0564517817837101\\
250	0.0494638898345657\\
300	0.0444587124442788\\
350	0.0405663718172311\\
};
\addlegendentry{\tabasco{}}

\addplot [color=mycolor1, line width=0.7pt,  mark size=1.6pt, mark=square, mark options={solid, mycolor1}]
  table[row sep=crcr]{%
50	0.149704371650435\\
100	0.0972037152391348\\
150	0.0756204852630571\\
200	0.0634377248096664\\
250	0.0553977761524816\\
300	0.0495430652326312\\
350	0.0450803758029725\\
};
\addlegendentry{ST-gaus}

\addplot [color=mycolor2, line width=0.7pt,  mark size=1.8pt,  mark=triangle, mark options={solid, rotate=90, mycolor2}]
  table[row sep=crcr]{%
50	0.164255042692645\\
100	0.102674039307579\\
150	0.078254684522151\\
200	0.0661806877499486\\
250	0.0577401065145807\\
300	0.0515156145373438\\
350	0.0467374906212217\\
};
\addlegendentry{ST-nong}

\addplot [color=red, line width=0.7pt,  mark size=1.8pt,   mark=x, mark options={solid, red}]
  table[row sep=crcr]{%
50	0.139613433374077\\
100	0.0899416756937583\\
150	0.0704959965837283\\
200	0.0598864438372418\\
250	0.0512122548101419\\
300	0.0465000995274802\\
350	0.0417521039677922\\
};
\addlegendentry{MnMx-Taper}

\end{axis}
\end{tikzpicture}%
}
\subfloat{% This file was created by matlab2tikz.
\definecolor{mycolor1}{rgb}{0.14200,0.67200,0.30300}%
\definecolor{mycolor2}{rgb}{0.54200,0.27200,0.60300}%
\begin{tikzpicture}

\begin{axis}[%
width=0.245\fwidth,
%height=0.33\fwidth,
at={(0\fwidth,0\fwidth)},
scale only axis,
xmin=50,
xmax=350,
 tick label style={font=\scriptsize} , 
xlabel style={font=\color{white!15!black}},
xlabel={$n$},
yticklabel=\pgfkeys{/pgf/number format/.cd,fixed,precision=2,zerofill}\pgfmathprintnumber{\tick},
ymin=0.0291609014728811,
ymax=0.144132577547404,
 ytick={0.04,0.06,0.08,0.1,0.12,0.14},
ylabel style={font=\color{white!15!black}},
axis background/.style={fill=white},
xmajorgrids,
ymajorgrids,
legend style={legend cell align=left, align=left, fill=none, draw=none}
]
\addplot [color=blue, line width=0.7pt,  mark size=1.8pt, mark=o, mark options={solid, blue}]
  table[row sep=crcr]{%
50	0.103327278941438\\
100	0.0655542939951947\\
150	0.0506863383132509\\
200	0.0422457227809808\\
250	0.0362714857515243\\
300	0.0321236350889386\\
350	0.0291609014728811\\
};
%\addlegendentry{Tabasco}

\addplot [color=mycolor1, line width=0.7pt,  mark size=1.6pt, mark=square, mark options={solid, mycolor1}]
  table[row sep=crcr]{%
50	0.124436282629773\\
100	0.0761802043580991\\
150	0.0576397545226367\\
200	0.0476491177800437\\
250	0.0409381530616623\\
300	0.0362527399144349\\
350	0.0327305855230006\\
};
%\addlegendentry{ST-gaus}

\addplot [color=mycolor2, line width=0.7pt,  mark size=1.8pt,  mark=triangle, mark options={solid, rotate=90, mycolor2}]
  table[row sep=crcr]{%
50	0.144132577547404\\
100	0.0848060270613279\\
150	0.0617372415229661\\
200	0.0506799338556683\\
250	0.0433436190266569\\
300	0.0386018354687574\\
350	0.0349287521500761\\
};
%\addlegendentry{ST-nong}

\addplot [color=red, line width=0.7pt,  mark size=1.8pt, mark=x, mark options={solid, red}]
  table[row sep=crcr]{%
50	0.112379990165522\\
100	0.0696886656047746\\
150	0.0524024179717632\\
200	0.0436565781591911\\
250	0.0371187586317442\\
300	0.0327553700648394\\
350	0.0300339950697892\\
};
%\addlegendentry{MnMx-Taper}

\end{axis}
\end{tikzpicture}%
}\hspace{2pt}
\subfloat{% This file was created by matlab2tikz.
\definecolor{mycolor1}{rgb}{0.14200,0.67200,0.30300}%
\definecolor{mycolor2}{rgb}{0.54200,0.27200,0.60300}%
\begin{tikzpicture}

\begin{axis}[%
width=0.245\fwidth,
%height=0.33\fwidth,
at={(0\fwidth,0\fwidth)},
scale only axis,
xmin=50,
xmax=350,
xlabel style={font=\color{white!15!black}},
 tick label style={font=\scriptsize} , 
xlabel={$n$},
ymin=0.07,
ymax = 0.6, 
ylabel style={font=\color{white!15!black}},
 ytick={0.1,0.2,0.3,0.4,0.5,0.6},
 yticklabel=\pgfkeys{/pgf/number format/.cd,fixed,precision=2,zerofill}\pgfmathprintnumber{\tick}, 
axis background/.style={fill=white},
xmajorgrids,
ymajorgrids,
legend style={legend cell align=left, align=left, fill=none, draw=none}
]
\addplot [color=blue,  line width=0.7pt, mark size=1.8pt, mark=o, mark options={solid, blue}]
  table[row sep=crcr]{%
50	0.218732227528755\\
100	0.149116649302869\\
150	0.11826635707236\\
200	0.0994372255299632\\
250	0.0893248999819819\\
300	0.079261191985478\\
350	0.0736911825828627\\
};
%\addlegendentry{Tabasco}

\addplot [color=mycolor1, line width=0.7pt, mark size=1.6pt,  mark=square, mark options={solid, mycolor1}]
  table[row sep=crcr]{%
50	3.34057236073248\\
100	1.64526447411866\\
150	1.00656624081186\\
200	0.736909638301744\\
250	0.678784886771282\\
300	0.526252368579283\\
350	0.477260815457979\\
};
%\addlegendentry{ST-gaus}

\addplot [color=mycolor2,  line width=0.7pt, mark size=1.8pt, mark=triangle, mark options={solid, rotate=90, mycolor2}]
  table[row sep=crcr]{%
50	0.339706913083853\\
100	0.192819506122445\\
150	0.14902980672626\\
200	0.121115115848432\\
250	0.108404983346109\\
300	0.0937070097141289\\
350	0.08796415392553\\
};
%\addlegendentry{ST-nong}

\addplot [color=red,  line width=0.7pt, mark size=1.8pt, mark=x, mark options={solid, red}]
  table[row sep=crcr]{%
50	0.357201945528367\\
100	0.217997105693702\\
150	0.164884355033958\\
200	0.138467072098963\\
250	0.12174441840549\\
300	0.108132392291832\\
350	0.0974023668896117\\
};
%\addlegendentry{MnMx-Taper}

\end{axis}
\end{tikzpicture}%
}
\subfloat{% This file was created by matlab2tikz.
\definecolor{mycolor1}{rgb}{0.14200,0.67200,0.30300}%
\begin{tikzpicture}

\begin{axis}[%
width=0.245\fwidth,
%height=0.33\fwidth,
at={(0\fwidth,0\fwidth)},
scale only axis,
xmin=50,
xmax=350,
xlabel style={font=\color{white!15!black}},
xlabel={$n$},
 tick label style={font=\scriptsize} , 
ymin=0.057144447665732,
ymax = 0.7, 
ylabel style={font=\color{white!15!black}},
axis background/.style={fill=white},
xmajorgrids,
yticklabel=\pgfkeys{/pgf/number format/.cd,fixed,precision=2,zerofill}\pgfmathprintnumber{\tick}, 
 ytick={0.1,0.2,0.3,0.4,0.5,0.6,0.7},
ymajorgrids,
legend style={legend cell align=left, align=left, fill=none, draw=none}
]
\addplot [color=blue, line width=0.7pt, mark size=1.8pt, mark=o, mark options={solid, blue}]
  table[row sep=crcr]{%
50	0.187554543961562\\
100	0.129034349187994\\
150	0.0971644159070281\\
200	0.0820806463464124\\
250	0.0700794293344638\\
300	0.0651676288909634\\
350	0.057144447665732\\
};
%\addlegendentry{Tabasco}

\addplot [color=mycolor1, line width=0.7pt, mark size=1.8pt, mark=square, mark options={solid, mycolor1}]
  table[row sep=crcr]{%
50	3.06240348037945\\
100	1.82042295647478\\
150	1.03634054499546\\
200	0.92227319060456\\
250	0.651398167178251\\
300	0.755384376732977\\
350	0.493365670732465\\
};
%\addlegendentry{ST-gaus}

\addplot [color=red, line width=0.7pt, mark size=2.0pt, mark=triangle, mark options={solid, rotate=180, red}]
  table[row sep=crcr]{%
50	0.273825424989417\\
100	0.188486748966457\\
150	0.120403605426103\\
200	0.114187216430875\\
250	0.0877826254753019\\
300	0.087584469595016\\
350	0.0749665201233768\\
};
%\addlegendentry{Tapering}

%\addplot [color=black, line width=0.7pt]
 % table[row sep=crcr]{%
%50	0.216783522760308\\
%100	0.143392307745548\\
%150	0.112376340006718\\
%200	0.0933902340326324\\
%250	0.0800297167752816\\
%300	0.0706885195141863\\
%350	0.0637888742909972\\
%};
%\addlegendentry{Tabasco - theory}

\end{axis}
\end{tikzpicture}%
}
\vspace{-0.3cm} 
\caption{Average NMSE curves  when samples are from a MVN distribution (upper panel)  and  MVT distribution with $\nu=5$ d.o.f.  (lower panel), $\M$ follows model 2 with $\alpha = 0.1$ (left panel) and $\alpha = 0.3$ (right panel), $p = 250$.} \label{fig:tapNMSE}
\end{figure}
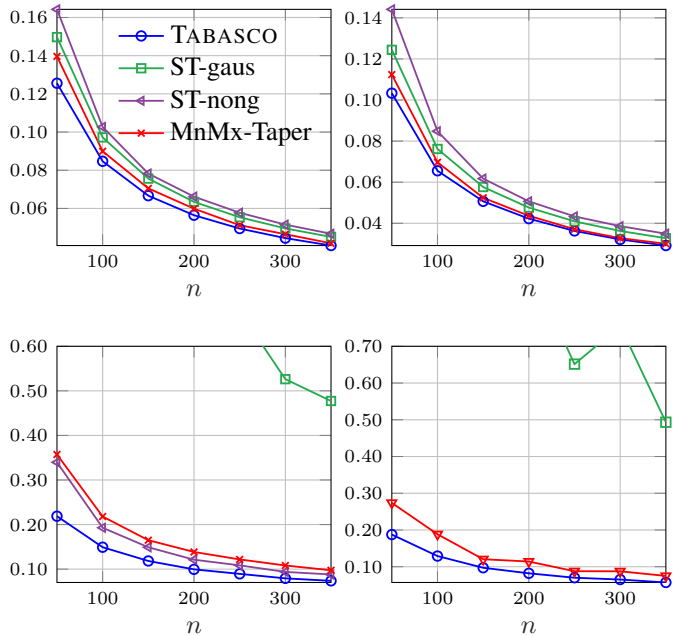

In {\bf Model~2} \cite{cai2010optimal}, $\M$ is defined by 
\beq\label{eq:Cai_model}
(\M)_{ij} =  
\begin{cases} 
1 &,  i=j  \\ 
\rho | i - j |^{-(\alpha +1 )} &, i \neq j, 
\end{cases}  
\eeq
where $\alpha$ is a decay parameter and $\rho$ is a correlation parameter. 
As in the study of \cite{cai2010optimal}, we set $\rho=0.6$, and \autoref{fig:Mij_models} illustrates the effect of decay parameter $\alpha$ in the case of $p=100$. 

\autoref{fig:tapNMSE} presents a comparison similar to \autoref{fig:AR_tdist} where we also included  the minimax risk tapering ({\bf MnMx-Taper}) estimator $\W(k^*) \circ \S$, where $k^*= \lfloor n^{1/(2(\alpha+1))} \rfloor$ is the optimal (oracle) bandwidth \cite[Section~6]{cai2010optimal}. The dimension is $p=250$. 
It should be noted that MnMx-Taper has  advantage over the other estimators since it uses the true decay parameter $\alpha$, which is unknown in practice.
\tabasco{} also uses tapering matrices $\W (k)$ as in \eqref{eq:Wminmax}, but ST-gaus and ST-nong are restricted to tapering matrices whose off-diagonal elements  are $0$-s or $1$-s. 
Hence, these are still computed with banding matrices $\W(k)$ as in \eqref{eq:Wband}. 
In either case, the optimal bandwidth $\hat k_o$ is chosen by consider the set of tapering matrices  
 $\mathbb{W} = \{ \bo W (k)  : k \in  [\![1,30]\!]\cup  [\![p-30,p]\!] \}$. 
As can be noted, \tabasco{} again outperforms other estimators for all values of $n$ and $\alpha$ and for both sampling distributions.  In the MVN case (top panel), \tabasco{} outperforms MnMx-Taper with a clear margin when $n$ is very small. This can be attributed to its ability to optimally shrink the tapered SCM towards a scaled  identity matrix when $n/p < 1$. However for $n \geq p$, \tabasco{} and MnMx-Taper estimator have similar performance, especially when $\alpha=0.3$. 

\begin{figure}[!t]
\centering
\setlength\fwidth{0.82\textwidth}
\subfloat{% This file was created by matlab2tikz.
\begin{tikzpicture}

\begin{axis}[%
width=0.23\fwidth,
%height=0.33\fwidth,
at={(0\fwidth,0\fwidth)},
scale only axis,
xmin=2,
xmax=18,
xlabel style={font=\color{white!15!black}},
xlabel={k},
ymin=0.0838069697070957,
ymax=0.11019657729508,
xlabel={bandwidth, $k$},
tick label style={font=\scriptsize} , 
ylabel style={font=\color{white!15!black}},
             yticklabel style={%
                 /pgf/number format/.cd,
                     fixed,
                     fixed zerofill,
                     precision=3,
                     },%ylabel={$\E[ \| \hat \M - \M \|_{\Fr}^2]$},
xtick={2,4,6,8,10,12,14,16,18},
 ytick={0.085,0.09,0.095,0.1,0.105,0.11},
axis background/.style={fill=white},
xmajorgrids,
ymajorgrids,
legend style={anchor=south west, legend cell align=left,  font = {\footnotesize}, align=right, draw=none, legend columns=2,at={(0.3,-0.47)}}
]
\addplot [color=blue,  line width=0.7pt,  mark size=1.8pt, mark=o, mark options={solid, blue}]
  table[row sep=crcr]{%
2	0.188530117638868\\
4	0.10271846613498\\
6	0.0860520840854068\\
8	0.0838069697070957\\
10	0.0865686209767171\\
12	0.0914654829731711\\
14	0.0973501364282572\\
16	0.103686178302792\\
18	0.11019657729508\\
20	0.11672871265758\\
22	0.123195111992565\\
24	0.129544717166694\\
26	0.13574785158656\\
28	0.141787869627492\\
30	0.147656289020973\\
32	0.153349840135404\\
34	0.158868619027783\\
36	0.164214900697289\\
38	0.169392360182408\\
40	0.17440555264593\\
42	0.179259561849643\\
44	0.183959760349047\\
46	0.188511645100083\\
48	0.192920724718847\\
50	0.197192442554498\\
52	0.201332124840055\\
54	0.205344946539041\\
56	0.20923590974666\\
58	0.213009831024783\\
60	0.216671335096225\\
62	0.220224853052726\\
64	0.223674623744623\\
66	0.227024697385685\\
68	0.230278940669114\\
70	0.233441042880651\\
72	0.236514522633332\\
74	0.239502734949888\\
76	0.242408878493767\\
78	0.245236002805072\\
80	0.247987015438937\\
82	0.250664688934384\\
84	0.253271667564553\\
86	0.255810473836116\\
88	0.258283514718326\\
90	0.260693087591386\\
92	0.263041385910795\\
94	0.265330504589168\\
96	0.267562445100724\\
98	0.269739120316076\\
100	0.2718623590767\\
102	0.273933910519548\\
104	0.275955448162936\\
106	0.277928573765148\\
108	0.279854820967253\\
110	0.281735658731517\\
112	0.283572494586564\\
114	0.28536667769005\\
116	0.287119501719248\\
118	0.288832207599498\\
120	0.290505986079973\\
122	0.292141980165775\\
124	0.293741287414878\\
126	0.295304962107965\\
128	0.296834017298746\\
130	0.298329426751924\\
132	0.299792126775511\\
134	0.301223017953815\\
136	0.302622966787048\\
138	0.303992807243087\\
140	0.305333342226636\\
142	0.306645344970656\\
144	0.307929560354676\\
146	0.309186706154251\\
148	0.310417474225625\\
150	0.31162253162934\\
152	0.312802521696356\\
154	0.313958065039977\\
156	0.315089760516687\\
158	0.316198186138825\\
160	0.317283899941808\\
162	0.318347440808475\\
164	0.319389329252935\\
166	0.320410068166205\\
168	0.321410143525707\\
170	0.322390025070652\\
172	0.323350166945137\\
174	0.324291008310736\\
176	0.325212973930209\\
178	0.326116474723884\\
180	0.327001908300161\\
182	0.327869659461515\\
184	0.328720100687266\\
186	0.329553592594346\\
188	0.330370484377193\\
190	0.331171114227845\\
192	0.331955809737244\\
194	0.332724888278715\\
196	0.333478657374502\\
198	0.334217415046229\\
200	0.334941450150054\\
202	0.335651042697316\\
204	0.33634646416134\\
206	0.337027977771111\\
208	0.337695838792431\\
210	0.338350294797163\\
212	0.338991585921145\\
214	0.33961994511129\\
216	0.340235598362393\\
218	0.340838764944126\\
220	0.341429657618669\\
222	0.34200848284942\\
224	0.342575441001177\\
226	0.343130726532195\\
228	0.343674528178469\\
230	0.344207029130609\\
232	0.344728407203613\\
234	0.345238834999879\\
236	0.345738480065726\\
238	0.346227505041725\\
240	0.346706067807095\\
242	0.34717432161843\\
244	0.347632415242979\\
246	0.348080493086738\\
248	0.348518695317537\\
250	0.348947157983356\\
};
\addlegendentry{\tabasco{}, $\hat \M_{\beta}$}

\addplot [color=red, dashed,  line width=0.7pt,  mark size=1.8pt,mark=square, mark options={solid, red}]
  table[row sep=crcr]{%
2	0.189955993927895\\
4	0.104431584105162\\
6	0.0896832615565728\\
8	0.0899027360016629\\
10	0.0955617089271843\\
12	0.103743727075036\\
14	0.113269542816193\\
16	0.123575205819153\\
18	0.134358592780255\\
20	0.145443947722059\\
22	0.156722409378353\\
24	0.168123133584293\\
26	0.179598134726811\\
28	0.191113817206337\\
30	0.202645996719681\\
32	0.214176845827406\\
34	0.225692951088712\\
36	0.237184038237307\\
38	0.248642112793763\\
40	0.26006086683522\\
42	0.271435260810703\\
44	0.282761223186829\\
46	0.294035431073784\\
48	0.305255147554295\\
50	0.316418099391858\\
52	0.327522383937772\\
54	0.338566397449649\\
56	0.349548779313354\\
58	0.360468368217346\\
60	0.371324167408205\\
62	0.382115316915699\\
64	0.392841071176991\\
66	0.403500780880033\\
68	0.414093878130954\\
70	0.424619864260136\\
72	0.43507829973805\\
74	0.445468795789111\\
76	0.455791007380992\\
78	0.46604462733458\\
80	0.476229381352299\\
82	0.486345023802815\\
84	0.496391334132065\\
86	0.506368113795261\\
88	0.51627518362428\\
90	0.526112381560415\\
92	0.535879560695137\\
94	0.545576587571296\\
96	0.555203340705635\\
98	0.564759709299835\\
100	0.574245592112832\\
102	0.583660896471473\\
104	0.593005537400257\\
106	0.602279436853913\\
108	0.611482523038942\\
110	0.620614729812451\\
112	0.629675996148272\\
114	0.638666265661792\\
116	0.647585486186127\\
118	0.656433609393391\\
120	0.665210590455542\\
122	0.673916387740124\\
124	0.682550962536841\\
126	0.691114278811353\\
128	0.699606302983252\\
130	0.70802700372551\\
132	0.716376351783015\\
134	0.724654319808131\\
136	0.732860882211479\\
138	0.740996015026329\\
140	0.749059695785155\\
142	0.757051903407165\\
144	0.764972618095637\\
146	0.77282182124416\\
148	0.780599495350802\\
150	0.788305623939536\\
152	0.795940191488144\\
154	0.803503183362091\\
156	0.810994585753639\\
158	0.818414385625941\\
160	0.825762570661471\\
162	0.833039129214531\\
164	0.840244050267304\\
166	0.84737732338943\\
168	0.854438938700436\\
170	0.861428886835122\\
172	0.868347158911351\\
174	0.875193746500314\\
176	0.881968641598816\\
178	0.888671836603604\\
180	0.895303324287486\\
182	0.901863097777105\\
184	0.908351150532249\\
186	0.914767476326556\\
188	0.92111206922956\\
190	0.92738492358988\\
192	0.933586034019563\\
194	0.939715395379431\\
196	0.945773002765391\\
198	0.951758851495609\\
200	0.957672937098499\\
202	0.963515255301493\\
204	0.96928580202047\\
206	0.974984573349872\\
208	0.980611565553398\\
210	0.986166775055295\\
212	0.991650198432126\\
214	0.99706183240508\\
216	1.00240167383269\\
218	1.00766971970401\\
220	1.01286596713215\\
222	1.01799041334824\\
224	1.02304305569568\\
226	1.02802389162471\\
228	1.0329329186874\\
230	1.03777013453272\\
232	1.04253553690207\\
234	1.04722912362494\\
236	1.05185089261486\\
238	1.05640084186551\\
240	1.06087896944712\\
242	1.06528527350299\\
244	1.06961975224622\\
246	1.07388240395662\\
248	1.07807322697775\\
250	1.08219221971414\\
};
%\addlegendentry{data2}

\end{axis}
\end{tikzpicture}%
}
\subfloat{% This file was created by matlab2tikz.
\begin{tikzpicture}

\begin{axis}[%
width=0.23\fwidth,
%height=0.33\fwidth,
at={(0\fwidth,0\fwidth)},
scale only axis,
xmin=2,
xmax=18,
xlabel style={font=\color{white!15!black}},
xlabel={bandwidth, $k$},
ymin=0.169603038849602,
ymax=0.237796104445795,
 tick label style={font=\scriptsize} , 
ylabel style={font=\color{white!15!black}},
xtick={2,4,6,8,10,12,14,16,18},
 ytick={0.17,0.19,0.21,0.23},
axis background/.style={fill=white},
xmajorgrids,
ymajorgrids,
legend style={anchor=south west, legend cell align=left,  font = {\footnotesize}, align=right, draw=none, legend columns=2,at={(-0.2,-0.47)}}
]

\addplot [color=red, dashed, line width=0.7pt,  mark size=1.8pt, mark=square, mark options={solid, red}]
  table[row sep=crcr]{%
2	0.252314442975713\\
4	0.194160501644641\\
6	0.205650972702826\\
8	0.23159831318618\\
10	0.262691115604645\\
12	0.296090732620476\\
14	0.33065121888011\\
16	0.36582490220224\\
18	0.401318437111787\\
20	0.436961168306168\\
22	0.472647377224081\\
24	0.508308249332154\\
26	0.543897161400158\\
28	0.579381462283172\\
30	0.614737640538039\\
32	0.649948359457984\\
34	0.685000570852554\\
36	0.719884277189121\\
38	0.754591696974678\\
40	0.789116688514135\\
42	0.82345434362127\\
44	0.857600695749555\\
46	0.891552506772088\\
48	0.925307108841565\\
50	0.958862285480451\\
52	0.992216181044058\\
54	1.02536723099308\\
56	1.05831410762518\\
58	1.0910556774269\\
60	1.12359096725624\\
62	1.15591913730345\\
64	1.18803945930375\\
66	1.21995129885501\\
68	1.25165410096975\\
70	1.28314737819527\\
72	1.31443070078742\\
74	1.34550368853742\\
76	1.37636600393804\\
78	1.40701734644106\\
80	1.43745744760936\\
82	1.46768606700572\\
84	1.49770298869202\\
86	1.52750801823598\\
88	1.55710098014235\\
90	1.58648171564013\\
92	1.61565008077025\\
94	1.64460594472701\\
96	1.67334918841546\\
98	1.70187970319264\\
100	1.73019738976616\\
102	1.75830215722769\\
104	1.78619392220269\\
106	1.81387260810052\\
108	1.84133814445136\\
110	1.86859046631854\\
112	1.89562951377666\\
114	1.92245523144695\\
116	1.94906756808279\\
118	1.97546647619934\\
120	2.00165191174171\\
122	2.02762383378732\\
124	2.0533822042784\\
126	2.07892698778098\\
128	2.1042581512676\\
130	2.12937566392103\\
132	2.15427949695655\\
134	2.17896962346098\\
136	2.20344601824651\\
138	2.22770865771792\\
140	2.25175751975164\\
142	2.27559258358553\\
144	2.29921382971832\\
146	2.32262123981772\\
148	2.34581479663631\\
150	2.3687944839345\\
152	2.39156028640981\\
154	2.41411218963215\\
156	2.43645017998396\\
158	2.4585742446055\\
160	2.48048437134421\\
162	2.50218054870817\\
164	2.52366276582287\\
166	2.54493101239169\\
168	2.56598527865888\\
170	2.58682555537574\\
172	2.60745183376885\\
174	2.627864105511\\
176	2.64806236269392\\
178	2.66804659780302\\
180	2.68781680369406\\
182	2.70737297357121\\
184	2.72671510096676\\
186	2.74584317972218\\
188	2.76475720397047\\
190	2.78345716811962\\
192	2.80194306683721\\
194	2.82021489503606\\
196	2.83827264786072\\
198	2.85611632067489\\
200	2.87374590904956\\
202	2.89116140875205\\
204	2.90836281573554\\
206	2.92535012612936\\
208	2.94212333622987\\
210	2.95868244249183\\
212	2.97502744152036\\
214	2.99115833006333\\
216	3.00707510500422\\
218	3.02277776335538\\
220	3.03826630225165\\
222	3.05354071894449\\
224	3.06860101079623\\
226	3.08344717527478\\
228	3.09807920994864\\
230	3.11249711248214\\
232	3.1267008806309\\
234	3.14069051223768\\
236	3.15446600522829\\
238	3.16802735760785\\
240	3.18137456745715\\
242	3.19450763292928\\
244	3.20742655224638\\
246	3.2201313236966\\
248	3.23262194563118\\
250	3.2448984164617\\
};
\addlegendentry{$\W \circ \S$}

\addplot [color=blue, line width=0.7pt,  mark size=1.8pt,mark=o,mark options={solid, blue}]
  table[row sep=crcr]{%
2	0.238854941657183\\
4	0.172773343394255\\
6	0.169603038849602\\
8	0.178298742201071\\
10	0.190156758968238\\
12	0.202678769452711\\
14	0.21497950478982\\
16	0.226724096729982\\
18	0.237796104445795\\
20	0.248173267717984\\
22	0.25787520324519\\
24	0.266939607698731\\
26	0.275410895966149\\
28	0.283334656257104\\
30	0.290754966076301\\
32	0.297713163572059\\
34	0.304247372957599\\
36	0.310392421049903\\
38	0.316179952087749\\
40	0.321638636694763\\
42	0.326794418452709\\
44	0.331670767651845\\
46	0.336288926311199\\
48	0.340668136705716\\
50	0.344825850196655\\
52	0.348777915680024\\
54	0.352538748314495\\
56	0.356121479862475\\
58	0.359538092264331\\
60	0.362799536134683\\
62	0.365915835820239\\
64	0.368896182548907\\
66	0.371749017063782\\
68	0.374482102992297\\
70	0.37710259206116\\
72	0.37961708213716\\
74	0.382031668954963\\
76	0.38435199228655\\
78	0.386583277212469\\
80	0.388730371072115\\
82	0.390797776597485\\
84	0.392789681671521\\
86	0.394709986096901\\
88	0.39656232571316\\
90	0.398350094158274\\
92	0.400076462534603\\
94	0.401744397207528\\
96	0.403356675937708\\
98	0.404915902523985\\
100	0.406424520113139\\
102	0.407884823314538\\
104	0.40929896924185\\
106	0.410668987590085\\
108	0.411996789844081\\
110	0.413284177703864\\
112	0.414532850802943\\
114	0.415744413787353\\
116	0.416920382815969\\
118	0.418062191536228\\
120	0.419171196583718\\
122	0.420248682649075\\
124	0.421295867151184\\
126	0.422313904551757\\
128	0.423303890342828\\
130	0.424266864735614\\
132	0.425203816076406\\
134	0.426115684012657\\
136	0.42700336243026\\
138	0.427867702180993\\
140	0.428709513617347\\
142	0.429529568950378\\
144	0.430328604444767\\
146	0.431107322464024\\
148	0.431866393377574\\
150	0.432606457340482\\
152	0.433328125955558\\
154	0.434031983826829\\
156	0.434718590012493\\
158	0.43538847938489\\
160	0.436042163904295\\
162	0.436680133812841\\
164	0.437302858754315\\
166	0.437910788825146\\
168	0.43850435556143\\
170	0.439083972866483\\
172	0.439650037883036\\
174	0.440202931813891\\
176	0.440743020694516\\
178	0.441270656120845\\
180	0.441786175935246\\
182	0.442289904873446\\
184	0.442782155174949\\
186	0.443263227159323\\
188	0.443733409770557\\
190	0.444192981091508\\
192	0.444642208830339\\
194	0.44508135078069\\
196	0.445510655257214\\
198	0.445930361507996\\
200	0.446340700105251\\
202	0.446741893315633\\
204	0.447134155451349\\
206	0.447517693203235\\
208	0.447892705956855\\
210	0.448259386092603\\
212	0.448617919270739\\
214	0.448968484702222\\
216	0.449311255406148\\
218	0.449646398454552\\
220	0.449974075205266\\
222	0.450294441523512\\
224	0.450607647992845\\
226	0.450913840116016\\
228	0.451213158506311\\
230	0.45150573906988\\
232	0.451791713179518\\
234	0.452071207840367\\
236	0.452344345847952\\
238	0.452611245938951\\
240	0.452872022935068\\
242	0.453126787880375\\
244	0.453375648172431\\
246	0.453618707687514\\
248	0.453856066900238\\
250	0.45408782299785\\
};
%\addlegendentry{$\hat \M_{\beta}$}

\end{axis}
\end{tikzpicture}%
}
\vspace{-0.25cm}
\caption{The true (theoretical) NMSE curves as a function of bandwidth $k$ for the tapered SCM $\W \circ \S$ and \tabasco{} $\hat \M_{\beta}$ when sampling from a MVN distribution (left panel) and MVT distribution (right panel) with $\nu=5$ d.o.f., 
 $\M$ follows model 2 with $\alpha = 0.1$, $n=100$ and $p=250$.} \label{fig:tapNMSE_vs_k}
\end{figure}
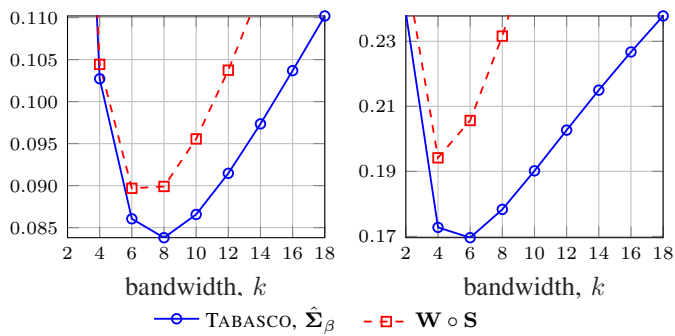

In the MVT case (lower panel of \autoref{fig:tapNMSE}), the performance differences are more clear. \tabasco{} outperforms MnMx-taper by a large margin. ST-gaus estimator completely fails due to the impulsive nature of the underlying sampling distributions.  
The results also illustrate that the performance of tapered SCM estimator is  dependent on the underlying sampling distribution more heavily than \tabasco{}. 
This is illustrated further in  \autoref{fig:tapNMSE_vs_k} where we compare the true theoretical NMSE curves of tapered SCM $\W \circ \S$ and \tabasco{} estimator $\hat \M_{\beta_0}$ as a function of bandwidth $k$ in the case where $n=100$ and when sampling from a MVN distribution (left panel) and MVT distribution (right panel) with $\nu=5$ d.o.f. following model 2 with $\alpha=0.1$. \autoref{fig:tapNMSE_vs_k} shows two important points. First, the performance differences between the tapered SCM and \tabasco{} are larger when the distribution is heavier tailed. This was evident already in \autoref{fig:tapNMSE}. Second, \tabasco{} with optimal bandwidth selection is able to estimate the optimal bandwidth rather accurately since the average (empirical) NMSE value seen in \autoref{fig:tapNMSE} at $n=100$ is close to the minimum true (theoretical) NMSE value.

%
% SECTION 8
% 

\section{Application to space-time adaptive processing}  \label{sec:applic}

Space time adaptive processing (STAP)  is a technique used in airborne phased array radar to detect moving target embedded in an interference background such as jamming or strong clutter \cite{Ward1998space}. 
The radar receiver consists in an array of $Q$ antenna elements processing $P$ pulses in a coherent processing interval. 
Within the tested sample $\mathbf{x}_0 \in \mathbb{C}^{p}$ with $p=P\cdot Q$, the received signal is composed of $i$) possible unknown targets responses; $ii$) unknown interferences (ground clutter) plus thermal noise. 
A detection problem for a given steering vector $\mathbf{p}$ is classically formalized as a binary hypothesis test: under $H_0$, $\mathbf{x}_0$ only contains the interference plus noise, or under $H_1$, $\mathbf{x}_0$ additionally contains a scaled observation of $\mathbf{p}$, i.e.:
\begin{align*}
\left\lbrace\begin{array}{lll}
H_0 : & \mathbf{x}_0 = \mathbf{n}_0 &;\; \mathbf{x}_i =  \mathbf{n}_i, \,\forall\, i\in [\![ 1,n ]\!]\\
H_1 : & \mathbf{x}_0 = \alpha \mathbf{p} + \mathbf{n}_0  &;\; \mathbf{x}_i =  \mathbf{n}_i, \,\forall\, i\in [\![ 1,n ]\!]
\end{array}\right.
\end{align*}
where $\x_i \in \mathbb{C}^p$, $i=1,\ldots,n$ is a secondary data set, assumed to contain i.i.d. and target-free realizations of the interference plus noise.
Usually, this disturbance $\mathbf{n}_i$ is modeled as centered complex Gaussian (or elliptically) distributed with covariance matrix $\mathbf{\Sigma}$.
In this context, efficient adaptive detection statistics can be built from the expression of the adaptive coherence estimator (ACE) detector \cite{Kraut2005adaptive}:
\begin{equation}\label{eq:ACE_detector}
\hat{\Lambda} ( \hat{\mathbf{\Sigma}} ) =
\frac{
| \mathbf{p}^\hop \hat{\mathbf{\Sigma}}^{-1} \mathbf{x}_0 |^2
}{
| \mathbf{p}^\hop \hat{\mathbf{\Sigma}}^{-1} \mathbf{p} |
| \mathbf{x}^\hop_0 \hat{\mathbf{\Sigma}}^{-1} \mathbf{x}_0 |
}  \mathop{\gtrless}_{H_0}^{H_1} \delta_{\hat{\mathbf{\Sigma}}} ,
\end{equation}
where $\hat{\mathbf{\Sigma}}$ is a \textit{plug-in} estimate of $\mathbf{\Sigma}$ computed from $\{\mathbf{x}_i\}_{i=1}^n$.
More specifically in STAP, the target $\mathbf{p}$ follows the steering vector model of \cite{Ward1998space}, which is function of the target angle of arrival (AoA) $\theta$ and velocity $v$.
The statistic \eqref{eq:ACE_detector} can thus be computed for a dictionary of steering vectors covering a 2D-grid on $\theta$ and $v$, yielding an adaptive detection map.

Using the SCM as estimate in \eqref{eq:ACE_detector} yields a generalized likelihood ratio test (GLRT) \cite{Kraut1999}, however, plug-in detectors can benefit from refined estimation processes in order to improve robustness, or to deal with limited sample support issues.
For example shrinkage to identity (also referred to as diagonal loading or robust beamforming \cite{li2003robust}) is a common procedure to improve several properties of the detector's output.
In the context of interference cancellation, tapering templates have been considered as a spectrum notch-widening technique \cite{guerci1999theory}, or to deal with modulation effects \cite{guerci2002principal}.

This section presents an experimental validation of \tabasco{} to illustrate the interest of both approach on real data. The STAP data is provided by the French agency DGA/MI: the clutter is real but the targets are synthetic.
The number of sensors is $Q = 4$ and the number of coherent pulses is $P = 64$, the size of the data is then $ p = QP = 256$. The center frequency and the bandwidth are respectively equal to $f_0 = 10 $GHz and the bandwidth $B = 5$MHz. The radar celerity is $V = 100 $m/s. The inter-element spacing is $d = 0,3$m and the pulse repetition frequency is $f_r = 1$kHz. The clutter to noise ratio is evaluated around $20$dB. 
We consider a test cell under $H_1$ with 10 targets of signal to clutter ratio around $-5$dB at various speed/angle and $n=397$ (all available) target-free secondary data to estimate the interference covariance matrix.
The tapering matrix is constructed as proposed in \cite{guerci1999theory}\footnote{The tapering in \cite{guerci1999theory} actually uses $\left[ \mathbf{T}_f  \right]_{ij} =  {\rm sinc}( (i-j) k / \pi )$ and $\left[ \mathbf{T}_\theta \right]_{ij} =  {\rm sinc}( (i-j) k / \pi )$, which performs a sliding window average on the estimated signal spectrum. The one considered here performs a linear combination of the original spectrum with such average. This modification was made so that the tapering matrix always conforms to the theoretical requirements $w_{ii}=1$ and $w_{ij} \geq 0$, but did not significantly impacted the output of the tested detectors.}, i.e. 
\begin{equation}
\begin{array}{l}
\mathbf{W}(k)=\mathbf{T}_f \otimes \mathbf{T}_\theta \\
\left[ \mathbf{T}_f  \right]_{ij} = (1+{\rm sinc}( (i-j) k / \pi ))/2 \in \mathbb{R}^{P\times P} \\
\left[ \mathbf{T}_\theta  \right]_{ij} = (1+{\rm sinc}( (i-j) k / \pi ))/2 \in \mathbb{R}^{Q\times Q}
\end{array}
\end{equation}
Note that index $k$ is here a "null width" parameter in $\mathbb{R}^+$ and not a bandwidth parameter in $[\![1,p]\!]$ as in \eqref{eq:Wband} or 
\eqref{eq:Wminmax}.

\begin{figure}
\centerline{\includegraphics[width=0.49\textwidth]{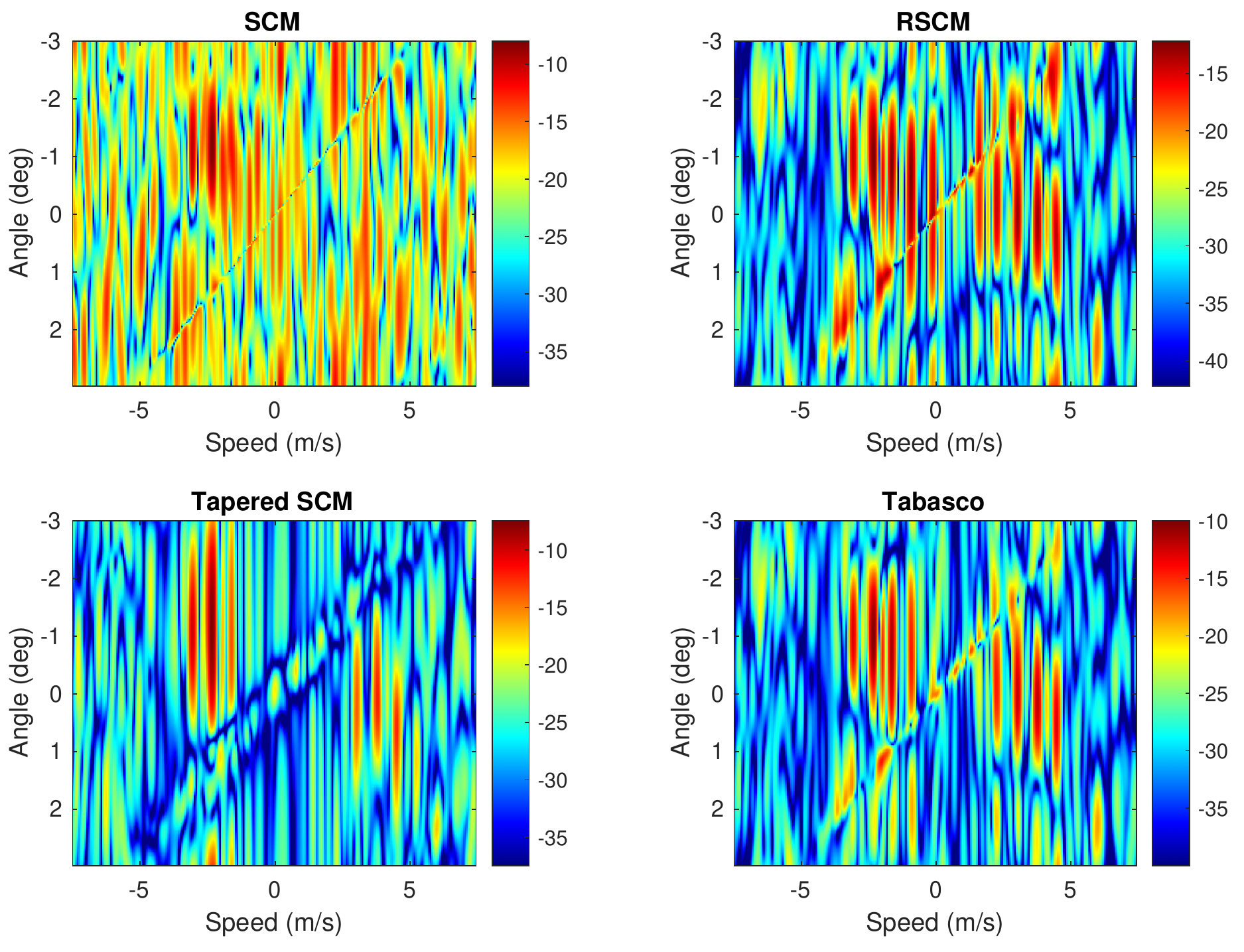}}
\vspace{-0.2cm} 
\caption{Output of various STAP detectors.}\label{fig:stap1}
\end{figure}

\autoref{fig:stap1} presents the detection map of $\hat{\Lambda}(\hat{\mathbf{\Sigma}})$ constructed with: 
$i$) the SCM; $ii$) the tapered SCM  $\W(k) \circ \S$ using bandwidth $k = 0.05$ (selected manually to obtain the best visual results); $iii$) \tabasco{} with the proposed adaptive selection of $\beta$ for $k=0$ (equivalent to RSCM, yielding $\beta = 0.9324$); $iv$) \tabasco{} with the proposed adaptive selection of $\beta$ and $k$ allowing $k\in[10^{-3},10^{-1}]$ (\tabasco{}, yielding $k= 0.0143$ and $\beta = 0.9929$).
First we can notice that the SCM provides an unreliable detection map, which is due to insufficient sample support in this configuration.
As observed in \cite{guerci1999theory} on another dataset, the covariance matrix tapering can widen the clutter notch (anti-diagonal of the detection map), which permits to clearly distinguish several targets.
However, this improvement is at the cost of canceling the response of slower targets (which are close to the canceled clutter ridge).
The shrinkage to identity of RSCM also greatly improves the detection process, as it allows us to detect the 10 targets, but still presents some false alarms on the clutter ridge.
Finally, \tabasco{} appears as an interesting trade-off by combining the two effects, and illustrates that the proposed NMSE-driven method still allows for a reasonable regularization parameters (both $\beta$ and $k$) selection in this detection application.

%
% SECTION 9
% 
 
\section{Conclusions and perspectives} \label{sec:concl}

We proposed \tabasco{}: a new covariance matrix estimator that jointly benefits from shrinkage to a scaled identity matrix and tapering of the SCM.
By assuming the samples to be generated from an unspecified ES distribution, we also derived an efficient and robust estimation method for the oracle regularization parameters that minimize the MSE.
Simulations studies illustrated that \tabasco{} outperforms existing regularized and tapered estimators in numerous setups.
Interestingly, if $\W = \mathbf{1}\mathbf{1}^\top$ belongs to the set of tapering matrices $\mathbb{W}$ considered,
the estimator can avoid applying tapering if this option does not provide reduction to the MSE.  
Thus \tabasco{} performs similarly to the regularized SCM proposed in \cite{ollila2019optimal} in this case, while significantly outperforming it when the tapering templates are valid. We also proposed two new novel estimators that measure the sphericity of the tapered covariance matrix.

%
% APPENDIX
% 

\appendix

\subsection{Proof  of  \autoref{th:beta0}} \label{app:th:beta0}  

Write $L(\be) = \MSE(\hat{\M}_{\be}) = \E[\| \hat{\M}_\be -\M\|_{\Fr}^2]$. Then note that
\begin{align}
L(\be)&= \E \left[  \left\| \be( \W \circ \S) + (1- \be)p^{-1} \tr(\S) \mathbf{I} - \M \right\|^2_{\Fr} \right] \notag  \\ 
&= \E \left[  \left \| \be(\W \circ \S - \M)  + (1- \be) \big ( p^{-1} \tr(\S)  \mathbf{I} - \M ) \right\|^2_{\Fr} \right] \notag  \\   
&= \be^2 a_1 +  (1- \be)^2  a_2 + 2\be(1-\be) a_3  \label{eq:Th1_Lbe}
\end{align}
where $a_3 =  \| \M \|_{\Fr}^2 - \| \bo V \circ \M \|_{\Fr}^2 +   \tilde a_3 $, and 
$a_1 =\mathrm{MSE}(\W \circ \S)$ is given in \eqref{eq:MSEtaper}, 
 \begin{align*}
a_2 &= \E \left [ \left\|  p^{-1} \tr(\S) \mathbf{I}  - \M  \right\|^2_{\Fr} \right] = \tilde a_3+  \| \M \|_{\Fr}^2 -  p \eta^2,  \\  
\tilde a_3 &= p^{-1} \E \left[  \tr \left( (\W \circ \S) - \M \right) \tr(\S) \right]  
= p^{-1}\E\big[	\tr (\S)^2  \big]  - \eta^2 p
\end{align*} 
and $\eta = \tr(\M)/p$. 
Note that $L(\be)$ is a convex quadratic function in $\be$ with a unique minimum given by 
\beq \label{eq:Th1_be0}
\be_o=  \frac{a_2-a_3}{ (a_1-a_3) + (a_2-a_3)}.  
\eeq 
Substituting the expressions for constants $a_1, a_2$ and $a_3$ into $\be_o $ yields the stated expressions in     \eqref{eq:beta0id}  and     \eqref{eq:beta0id1}. 
 In this regard, it is useful to notice that  $a_2-a_3 = \| \bo V \circ \M \|_{\Fr}^2  - p \eta^2 = \|  \bo V \circ \M  - \eta \bo I \|_{\Fr}^2 = p(\gamma_{\V} -1) \eta^2$. 
Expression     \eqref{eq:beta0id2} can be deduced from   \eqref{eq:beta0id1} by  using \eqref{eq:MSEtaper} and  then simplifying the expression. 

The expression for MSE of $\hat \M_{\be_o}$ follows by substituting $\be_o$ into expression for $L(\be)$ in \eqref{eq:Th1_Lbe} 
and using the relation, $(1-\be_o)  (a_2-a_3) = \be_o(a_1-a_3)$,  
which follows from \eqref{eq:Th1_be0}. This gives
\[
L(\be_o) = a_2 - \be_o(a_2-a_3)  =  a_3 + (1- \be_o) (a_2-a_3).
\]
This gives the stated MSE expression after noting that  $a_2-a_3 = \left\| \bo V \circ\M - \eta \mathbf{I} \right\|_{\Fr}^2 $.

\subsection{Proof  of  \autoref{lem:EtrTaper}} \label{app:lem:EtrTaper}

Before proceeding with the proof we introduce some definitions and  results that are used in the sequel. 
First, we let  $\commat$ denote the  $p^2 \times p^2$ commutation matrix defined as 
a  block matrix whose  $ij$th block is equal to a $p\times p$ matrix that has a $1$ at
element $ji$ and zeros elsewhere, i.e., $\commat = \sum_{i,j} \e_i \e_j^\top
\otimes \e_j \e_i^\top$. It also has the following important
properties~\cite{magnus_neudecker:1999}: $\commat \ve(\mathbf{A})= \ve(\mathbf{A}^\top)$ and
$\commat (\mathbf{A} \otimes \mathbf{B}) \commat = (\mathbf{B} \otimes \mathbf{A})$ for any $p \times
p$ matrices $\mathbf{A}$ and $\mathbf{B}$, where 
$\ve(\A)$ vectorizes matrix $\A$ by stacking the columns of the matrix on top of each other. 
We then have the following identities.

 \begin{lemma}  \label{lem:EtrTaper_ident} The following holds:
\begin{enumerate}[a)] 
\item \label{lem:EtrTaper_identA} $\| \A \circ \B \|_{\Fr}^2 = \tr\left( \ve(\A) \ve(\A)^\top \circ \ve(\B) \ve(\B)^\top \right) $ for all $\A, \B \in \R^{m \times n}$. 
\item \label{lem:EtrTaper_identB}  $ \| \A \circ \B \|_{\Fr}^2 = \tr \left( \ve(\A) \ve(\A)^\top  \circ   \commat (\B \otimes \B) \right) $ 
$\forall \A \in \R^{m \times m}$ and $\forall \B \in \Sym{m}$. 
\item \label{lem:EtrTaper_identC} $\d^\top_{\B} (\A \circ \A) \d_{\B}   = \tr \left(  \ve(\A)  \ve(\A)^\top  \circ   (\B \otimes \B) \right) $ $\forall   \A, \B \in \R^{m\times m}$. 
\item $\tr \big( (\D_{\B} \A)^2 \big)  = \d^\top_{\B} (\A \circ \A) \d_{\B} $ for 
all   $\A \in \Sym{m}$ and $\B \in \R^{m \times m}$.  
\end{enumerate} 
\end{lemma} 

\begin{proof} Let $\A=(a_{ij})$ and $\B = (b_{ij})$. 
{\it a)} First note that 
\begin{align*}
\| \A \circ \B \|_{\Fr}^2 &= \tr\left( \ve(\A \circ \B) \ve(\A \circ \B)^\top \right) \\ &= \tr\left( \ve(\A) \ve(\A)^\top \circ \ve(\B) \ve(\B)^\top \right). 
\end{align*} 
 {\it b)}  It is a simple matter to verify that for all $\B \in \Sym{m}$ it holds that 
 $\diag(\ve(\B) \ve(\B)^\top)=\diag(\commat ( \B \otimes \B))$. Thus 
\begin{align*}
 &\tr \left( \ve(\A) \ve(\A)^\top  \circ   \commat (\B \otimes \B) \right)   \\ 
 &\qquad=  \tr \left( \ve(\A) \ve(\A)^\top  \circ  \ve(\B) \ve(\B)^\top \right)
 \end{align*} 
 which gives the stated result due to a)-part. 
  {\it c)}  It is a simple task to verify that  
the trace of the Hadamard product of $\ve(\A) \ve(\A)^\top $ with  $\B \otimes \B$ equals $ \sum_{i,j} b_{ii} a_{ij}^2 b_{jj}$ which is  equivalent with $\d^\top_{\B} (\A \circ \A) \d_{\B}$. {\it d)} Follows from \cite[Lemma~7.5.2]{horn2012matrix}.   
\end{proof} 

Write $\w= \ve(\W)$. Using  \autoref{lem:EtrTaper_ident}\ref{lem:EtrTaper_identA} we first notice that 
\beq \label{eq:Expec_norm_WoS}
\E \left[\| \W \circ \S \|_{\Fr}^2 \right]  = \tr\left( \w\w^\top \circ \E[ \ve(\S) \ve(\S)^\top ] \right). 
\eeq 
We then recall that the (variance-)covariance matrix of $\S$  when sampling from an elliptical population $\mathcal E_p(\bmu, \M,g)$ is given by \cite[Theorem~2]{ollila2019optimal}: 
\begin{align}
\cov&( \ve(\S) ) =  \E[ \ve(\S)\ve(\S)^\top ]  - \ve(\M) \ve(\M)^\top  \label{eq:varvecS_apu}   \\
&=  \tau_1 ( \mathbf{I} + \commat) (\M \otimes \M) + \tau_2 \ve(\M) \ve(\M)^\top ,  \label{eq:varvecS} 
\end{align} 
where $\tau_1$ and $\tau_2$ are constants defined in \eqref{eq:tau_1and2}. Equations \eqref{eq:varvecS_apu} and \eqref{eq:varvecS}  then imply that 
\begin{align} \label{eq:Expec_veSveSt}
\E[ &\ve(\S)\ve(\S)^\top ]  =   \cov(  \ve(\S) )  + \ve(\M) \ve(\M)^\top  \notag  \\ 
&= \tau_1 ( \mathbf{I} + \commat) (\M \otimes \M)  + (1+\tau_2) \ve(\M) \ve(\M)^\top . 
\end{align} 
 Inserting \eqref{eq:Expec_veSveSt} into \eqref{eq:Expec_norm_WoS} yields
\begin{align*} 
\E \left[\| \W \circ \S \|_{\Fr}^2 \right]    
&= (1+\tau_1+\tau_2)  \| \W \circ \M \|_{\Fr}^2 + \tau_1 \tr \big( (\D_{\M} \W)^2 \big).
\end{align*} 
simply by invoking identities in \autoref{lem:EtrTaper_ident}.  This proves the first identity. 

Next we note that 
\beq \label{eq:Expec_D_S_W_sq}
\E \big[ \tr( ( \D_{\S} \W)^2 ) \big]  = \sum_{i,j=1}^p \E[ s_{ii} s_{jj} ] w_{ij}^2.
\eeq 
Equation  \eqref{eq:Expec_veSveSt} implies that 
\beq \label{eq:Expec_Sii_Sjj}
\E[s_{ii}s_{jj}] =  2 \tau_1 \sigma_{ij}^2 + (1+\tau_2) \sigma_{i}^2 \sigma_{j}^2. 
\eeq 
Thus inserting \eqref{eq:Expec_Sii_Sjj} into \eqref{eq:Expec_D_S_W_sq} yields 
\begin{align*} 
\E \big[ \tr( ( \D_{\S} \W)^2 ) \big]  &= 2 \tau_1 \sum_{i,j=1}^p  w_{ij}^2 \sigma_{ij}^2 +(1+\tau_2)  \sum_{i,j=1}^p w_{ij}^2 \sigma_{i}^2 \sigma_{j}^2 \\ 
&=2 \tau_1 \| \W \circ \M \|_{\Fr}^2 + (1+\tau_2)  \Tr((\D_{\M} \W)^2 )
\end{align*} 
which proves the latter claim. 

\subsection{Proof  of  \autoref{th:Ell1}} \label{app:th:Ell1}

Let us express the SSCM as 
\[
\SSCMshape = \frac{1}{n} \sum_{i=1}^n \v_i \v_i^\top, \quad \mbox{where } \, \mathbf{v}_i = \sqrt{p} \frac{\x_i}{\| \x_i \|}.  
\] 
Hence 
\begin{align*}
&\frac{\|  \bo W \circ  \hat \La \|_{\Fr}^2}{p} 
= \frac{1}{pn^2}   \tr \Big(  \big(\W  \circ \v_1 \v_1^\top +  \cdots +   \W  \circ \v_n \v_n^\top \big)^2 \Big)  \notag \\ 
&= \sum_{i=1}^n  \frac{\| \W  \circ \v_i \v_i^\top \|_{\Fr}^2}{pn^2}  
  +   \sum_{i \neq j}^n \frac{  \tr \big( \big(\W  \circ \v_i \v_i^\top \big)  \big(\W  \circ \v_j \v_j^\top \big)  \big) }{pn^2}  . 
\end{align*}
Then since $\v_i$-s are i.i.d.,  and $\E[\SSCMshape] =  \E[ \v_i \v_i^\top]$ for all $i$,  the expectation of the 2nd term  is 
\begin{align*} 
\sum_{i \neq j}  \frac{ \E\big[  \tr\! \big( (\W  \circ \v_i \v_i^\top ) (\W  \circ \v_j \v_j^\top) \big) \big] }{pn^2}
 &= \frac{n-1}{n}   \frac{\|\W \circ \La_{\textup{sgn}} \|_{\Fr}^2}{p}, 
\end{align*}
where $\La_{\textup{sgn}}= \E[\SSCMshape ]$. 
The expectation of the 1st terms is 
\begin{align*} 
 \sum_{i}  \frac{ \E \big[ \| \W  \circ \v_i \v_i^\top \|_{\Fr}^2 \big]}{pn^2} &=  \frac{ \E \big[ \| \W  \circ \v \v^\top \|_{\Fr}^2 \big]}{pn}  \\ 
   &=  \frac{\E\Big[  \d^\top (\W \circ \W) \d \Big] }{pn}  
  \end{align*} 
 where $\d= (v_1^2,\ldots,v_p^2)^\top$ contains the diagonal elements of $\v \v^\top$, where $\v =_d \v_i$ and $=_d$ reads ``has the same distribution as''.  Furthermore, write 
 $\D = \diag(\v \v^\top)$. 
 Thus we have that 
\begin{align} 
&\frac{n}{n-1} \cdot \frac{ \E \big[ \|  \bo W \circ  \hat \La \|_{\Fr}^2 \big] }{p} \notag   \\ 
&=  \frac{\| \W \circ \La_{\textup{sgn}} \|^2_{\Fr} }{p}+ \frac{\E\big[  \d^\top (\W \circ \W) \d \big] }{p(n-1)}  \label{eq:ell1_proof1}. 
 \end{align} 
Next note that $\D_{\hat \La} = \diag(\hat \La)$ can be  written as 
\[
\D_{\hat \La} =  \frac{1}{n} \big( \D_{1}  + \ldots +  \D_{n} \big) ,
\] 
where $\D_i = \diag(\v_i \v_i^\top)$. Furthermore, let $\d_i = (v_{i1}^2,\ldots,v_{ip}^2)^\top$ denote a random vector containing the diagonal elements of $\v_i \v_i^\top$. Then we get 
\begin{align*} 
&\tr \Big ( \big(\D_{\hat \Lambda} \W \big)^2 \Big) = \frac{1}{n^2} \tr \Big ( \Big( \D_1 \W + \ldots + \D_n \W \Big)^2 \Big) \\ 
&= \frac{1}{n^2} \sum_{i=1}^n  \tr \Big ( \Big( \D_i \W \Big)^2 \Big)  +  \frac{1}{n^2} \sum_{i \neq j } \tr \Big ( ( \D_i \W) \D_j \W \Big) \\ 
&= \frac{1}{n^2} \sum_{i=1}^n  \d_i (\W \circ \W) \d_i +  \frac{1}{n^2} \sum_{i \neq j } \tr \Big ( ( \D_i \W) \D_j \W \Big) . 
 \end{align*} 
 Thus 
 \begin{align} 
 &\frac{1}{p(n-1)} \E \Big[ \tr \Big ( \big(\D_{\hat \Lambda} \W \big)^2 \Big) \Big] \notag \\ 
 &=  \frac{\E[ \d(\W \circ \W) \d]}{pn(n-1)} + \frac{1}{pn}  \tr \Big( \big( \E[ \D ] \W \big)^2 \Big) \notag  \\
 &=  \frac{\E[ \d(\W \circ \W) \d]}{pn(n-1)} + \frac{1}{pn}  \E[ \d ]^\top (\W \circ \W) \E[\d] . \label{eq:ell1_proof2}
 \end{align} 
 Using \eqref{eq:ell1_proof1} and \eqref{eq:ell1_proof2} we then obtain that 
 \beq \label{th:Ell1_apueq1}
 \E[ \hat \gamma_{\W} ] =  \frac{\| \W \circ \La_{\textup{sgn}} \|^2_{\Fr}}{p}+ \frac{1}{n} \varepsilon , 
 \eeq
 where 
 \begin{align} 
 \varepsilon &=  \frac{1}{p} \Big(  \E[ \d(\W \circ \W) \d] -  \E[ \d ]^\top (\W \circ \W) \E[\d]  \Big)  \notag \\ 
  &=\frac{1}{p} \Big( \sum_{i=1}^p  \var(v_i^2) + \sum_{i\neq j}^p  w_{ij} \cov(v_i v_j)  \Big) \to 0  \quad  \mbox{as $p \to \infty$} .  
\label{th:Ell1_apueq2}
 \end{align} 
Next note that  $ \La_{\textup{sgn}} = \E[\SSCMshape]  = \La + o (\|  \La \|_{\Fr})$ when (A) holds  by \cite[Theorem~2]{raninen2020linear}
This  fact together with \eqref{th:Ell1_apueq1}  and \eqref{th:Ell1_apueq2} imply that 
\[ 
 \E[ \hat \gamma_{\W} ] \to    \frac{\|\W \circ \La \|^2_{\Fr}}{p} = \gamma_{\W}   
 \] 
 as $p \to \infty$ under assumption (A). Thus we have proven the claim.

\subsection{Proof  of  \autoref{lem:EtrTaper_c}: complex case} \label{app:lem:EtrTaper_c}

In our proof we will use the following identities. 
 \begin{lemma}  \label{lem:EtrTaper_ident_c} The following holds:
\begin{enumerate}[a)] 
\item \label{lem:EtrTaper_identA_c} $\| \A \circ \B \|_{\Fr}^2 = \tr\left( \ve(\A) \ve(\A)^\hop \circ \ve(\B) \ve(\B)^\hop \right) $ for all $\A, \B \in \C^{m \times n}$. 
\item \label{lem:EtrTaper_identC_c} $\d^\top_{\B} (\A \circ \A) \d_{\B}  
= \tr \left(  \ve(\A)  \ve(\A)^\top  \circ   (\B^* \otimes \B) \right) $ $\forall   \A \in \C^{m\times m}$ and 
$ \B \in \SymC{m}$.
\item $\tr \big( (\D_{\B} \A)^2 \big)  = \d^\top_{\B} (\A \circ \A) \d_{\B} $ for 
all   $\A \in \Sym{m}$ and $\B \in \C^{m \times m}$. 
\end{enumerate} 
\end{lemma} 

\begin{proof} {\it a,b)} proofs of the identities are as proofs of \autoref{lem:EtrTaper_ident}a),b). {\it c}) follows directly from \cite[Lemma~7.5.2]{horn2012matrix}.   \end{proof}

Write $\w= \ve(\W)$. Using  \autoref{lem:EtrTaper_ident_c}\ref{lem:EtrTaper_identA_c} we first notice that 
\beq \label{eq:Expec_norm_WoS_complex}
\E \left[\| \W \circ \S \|_{\Fr}^2 \right]  = \tr\left( \w\w^\top \circ \E[ \ve(\S) \ve(\S)^\hop ] \right). 
\eeq 
We then recall that the (variance-)covariance matrix of $\S$  when sampling from a complex elliptically symmetric distribution  $\mathbb{C} \mathcal E_p(\bmu, \M,g)$ is \cite[Theorem~3]{raninen2021variability}:
\begin{align}
\cov&( \ve(\S) ) =  \E[ \ve(\S)\ve(\S)^\hop ]  - \ve(\M) \ve(\M)^\hop  \label{eq:varvecS_apu_c}   \\
&=  \tau_1  (\M^* \otimes \M) + \tau_2 \ve(\M) \ve(\M)^\hop ,  \label{eq:varvecS_c} 
\end{align} 
where $\tau_1$ and $\tau_2$ are constants defined in \eqref{eq:tau_1and2}. Equations \eqref{eq:varvecS_apu_c} and \eqref{eq:varvecS_c}  then imply that 
\begin{align} \label{eq:Expec_veSveSt_c}
\E[ &\ve(\S)\ve(\S)^\hop ]  =   \cov(  \ve(\S) )  + \ve(\M) \ve(\M)^\hop  \notag  \\ 
&= \tau_1 (\M^* \otimes \M)  + (1+\tau_2) \ve(\M) \ve(\M)^\hop . 
\end{align} 
 Inserting \eqref{eq:Expec_veSveSt_c} into \eqref{eq:Expec_norm_WoS_complex} yields
\begin{align*} 
\E \left[\| \W \circ \S \|_{\Fr}^2 \right]    
&= (1+\tau_2)  \| \W \circ \M \|_{\Fr}^2 + \tau_1 \tr \big( (\D_{\M} \W)^2 \big).
\end{align*} 
simply by invoking identities in \autoref{lem:EtrTaper_ident_c}.  This proves the first identity.  The proof of latter part $\E \big[ \tr( ( \D_{\S} \W)^2 ) \big] $ is as earlier in the real-valued case in \autoref{app:lem:EtrTaper}.

%\section*{Acknowledgment}
%The authors would like to thank the reviewers for insightful comments which
%helped to improve the paper.

\ifCLASSOPTIONcaptionsoff
\newpage
\fi

\end{document}